\documentclass[12pt,letterpaper]{article}

\usepackage{amsmath}
\usepackage{amsfonts}
\usepackage{amssymb}
\usepackage{epsf}
\usepackage{epsfig}
\usepackage[numbers,sort&compress]{natbib}

\usepackage[hang,small,bf]{caption}

\DeclareRobustCommand{\SkipTocEntry}[4]{}

\def\beq{\begin{equation}}
\def\eeq{\end{equation}}
\def\bea{\begin{eqnarray}}
\def\eea{\end{eqnarray}}

\newcommand{\hii}{${\rm HII\ }$}

\newcommand{\bc}{\begin{center}}
\newcommand{\ec}{\end{center}} 

\usepackage{hyperref}
\providecommand{\eprint}[1]{\href{http://arxiv.org/abs/#1}{#1}}
\providecommand{\adsurl}[1]{\href{#1}{ADS}}

\def\n{{\bf  \hat n}}

\newcommand{\mpch}{\rm Mpc/h}

\newcommand{\bma}{\begin{math}}
\newcommand{\ema}{\end{math}}
\newcommand{\bit}{\begin{itemize}}
\newcommand{\eit}{\end{itemize}}

\def\apj{ApJ}

\def\apjl{ApJ Lett.}
\def\mnras{MNRAS}
\def\prd{Phys. Rev. D}
\def\aj{AJ}

\def\aap{A\& A}
\def\pasj{Pub. Astron. Soc. Japan}

\begin{document}

\vspace{5mm} \vspace{0.5cm}

\begin{center}

{\large CMBPol Mission Concept Study} \\
\vskip 15pt {\Large Reionization Science with the Cosmic Microwave Background }
\\[1.0cm]


{Matias Zaldarriaga$^{\rm 1, 2}$, Loris Colombo$^{3}$, Eiichiro Komatsu$^{4}$, Adam Lidz$^{1}$, Michael Mortonson$^{5}$, S. Peng Oh$^{6}$, Elena Pierpaoli$^{3}$, Licia Verde$^{7,8}$, Oliver Zahn$^{9,10}$}
\\[0.5cm]

\end{center}

\vspace{2cm} \hrule \vspace{0.3cm}
{\small  \noindent \textbf{Abstract} \\[0.3cm]
\noindent We summarize existing  constraints on the epoch of reionization and discuss the observational probes that are sensitive to  the process. We focus on the role large scale polarization can play. Polarization probes the integrated optical depth across the entire epoch of reionization. Future missions such as Planck and CMBPol will greatly enhance our knowledge of the reionization history,  allowing us to measure the time evolution of the ionization fraction.  As large scale polarization probes  high redshift activity, it can best constrain models where the Universe was fully or partially ionized at early times. In fact, large scale polarization could be our only probe of the highest redshifts.

 \vspace{0.5cm}  \hrule
\def\thefootnote{\arabic{footnote}}
\setcounter{footnote}{0}

\vspace{1.0cm}


\begin{center}


{\small \textit{$^{\rm 1}$ Center for Astrophysics, Harvard University, Cambridge, MA 02138, USA}}

{\small \textit{$^{\rm 2}$ Department of Physics, Harvard University, Cambridge, MA 02138, USA}}

{\small \textit{$^{\rm 3}$ Department of Physics and Astronomy, University of Southern California, Los Angeles, CA 90089, USA }}

{\small \textit{$^{\rm 4}$ Department of Astronomy, The University of Texas at Austin, Austin, TX 78712, USA}}

{\small \textit{$^{\rm 5}$ Department of Physics, University of Chicago, Chicago, IL 60637, USA}}

{\small \textit{$^{\rm 6}$ Department of Physics, University of California, Santa Barbara, CA 93106, USA}}

{\small \textit{$^{\rm 7}$ ICREA \& Institute of Space Sciences (CSIC-IEEC), Campus UAB, Bellaterra, Spain}}

{\small \textit{$^{\rm 8}$ Department of Astrophysical Sciences, Princeton, NJ 08540, USA}}

{\small \textit{$^{\rm 9}$ Berkeley Center for Cosmological Physics, University of California, Berkeley, Berkeley, CA 94720, USA}}

{\small \textit{$^{\rm 10}$ Lawrence Berkeley National Laboratory, Berkeley, CA 94720, USA}}

\end{center}

\newpage
\tableofcontents


 \newpage
\section{The Epoch of Reionization: How the Universe emerges from its dark ages}

\subsection{Introduction}


Perhaps the most important goal of observational cosmology  is to understand how structure arose
in our Universe, how the stars, galaxies, and quasars 
around us today and at high redshifts formed.  We currently believe that 
small density perturbations were imprinted onto the Universe
during the inflationary era which then grew through gravitational
instability, producing halos as well as the cosmic web of sheets and
filaments.  Measurements of the cosmic microwave background
(CMB) anisotropies have been instrumental in establishing this picture 
(e.g. \cite{Komatsu08}).

To understand the CMB measurements or the large scale clustering properties of galaxies one only needs to understand the structure formation process in the linear regime. However, in order to understand how baryons
collapsed into bound objects such as galaxies and galaxy clusters, and
to determine how these objects affect their surroundings we need to 
take structure formation beyond the well-understood linear regime.  At low or
moderate redshifts ($z < 6$), galaxies and quasars can be observed
directly with existing technology.  However, the first
generations of protogalaxies are not yet similarly accessible and the
full population of galaxies and quasars at intermediate redshifts
($z\sim 3$) remains unknown.

One approach to investigating the properties of galaxies and quasars
is through their impact on the intergalactic medium (IGM).  In
particular, it is believed that radiation from these objects reionized
the IGM and that this process depends on the distribution and nature
of the ionizing sources.  Thus, an understanding of reionization
coupled with detailed measurements has the potential to significantly
constrain models for the formation and evolution of galaxies and
quasars.

Observational evidence indicates that the IGM was reionized in two
distinct phases.  The first phase, completed by $z\approx 6$, led to the
ionization of hydrogen as well as singly ionizing helium.  However,
the early sources of ionizing radiation would not have been energetic
enough to doubly ionize helium (if they were stellar).  The second
phase of the reionization of the IGM, corresponding to the conversion
of HeII into HeIII, thus likely occurred later, when a sufficient
population of hard sources (i.e. quasars) developed, and observations
indicate that this happened at $z\approx 3$.

Reionization encodes a wealth of information about
the ionizing sources.  For example, in the case of hydrogen
reionization, the first sources of radiation likely formed
in overdense regions, ionizing pockets of nearby gas.  As more
sources formed these
\hii regions grew
 and eventually overlapped.  Because the structure of these \hii
regions and their growth depend on the nature, clustering, and
abundance of the sources, the timing, morphology, and duration of
reionization are highly sensitive to the properties of the first
luminous objects and their relationship to the IGM (e.g.,
\cite{wyithe03,cen03,haiman03,mackey03,yoshida03-semian,ybh,
sok03,sok04,yoshida06}).

Empirical studies of the reionization of hydrogen (sometimes referred
to as the Epoch of Reionization [EoR]) have placed constraints on this
event but have not yet mapped it in detail. One constraint comes from 
the effects of the ionized gas on the
CMB.  The free electrons Thomson scatter CMB photons, washing out the
intrinsic anisotropies but generating a polarization signal on large
angular scales.  The total scattering optical depth $\tau_{\rm es}$ is
proportional to the column density of ionized hydrogen, so it provides
an integral constraint on the reionization history.  Recently, the
{Wilkinson Microwave Anisotropy Probe}\footnote{See
http://map.gsfc.nasa.gov/.}  (WMAP) collaboration used five
years worth of data to improve their determination of $\tau_{\rm es}$
\cite{Komatsu08}.  The combined analysis of WMAP and large scale
structure data such as that coming from the Sloan Digital Sky 
Survey\footnote{See http://www.sdss.org/.} (SDSS) indicates that
reionization probably happened in the redshift range $z_r =
11\pm 1.4$.   More detailed information
on the reionization history could be obtained by precisely measuring
the large angular scale polarization \cite{zal97,kaplinghat03,hu03}. 
It is the goal of this white paper to discuss this possibility further. 

Along with the large scale CMB polarization measurements, we anticipate a wealth
of other upcoming observations that will increase our knowledge of the EoR.
The EoR and the preceding
`dark ages' -- i.e., the period after cosmological recombination but
before the birth of the first luminous sources -- are the only
remaining phases of structure formation as yet to be directly
observed. As such, the study of the EoR is a research frontier for both observational and
theoretical  cosmology, and will grow as an increasingly active and
exciting research field in upcoming years.   
We describe current constraints and future prospects from other
observations in \S \ref{sec:obs_probes}, and consider the way in which large scale CMB
polarization measurements may complement these observations throughout this white paper.

\subsection{Motivating questions and how to answer them}
\label{motq}

We believe that the study of reionization will shed light on the early
process of structure formation. In practice we want to answer several
specific questions about this early time in the history of our
Universe.   We want to determine which sources were
responsible for reionization. Were they a population of early galaxies
similar to present day ones? Were Population III stars important for
this process? Did quasars or mini-quasars play a role? Did the sources
of ionizing radiation sit in high or low mass galaxies? Did the
ionizing radiation produced by the first sources have a strong
feedback effect on the galaxy formation process, drastically changing
the way galaxy formation proceeded thereafter?


In some sense the study of reionization is akin to using the IGM as a
laboratory for learning about the first sources and the earliest
stages of the structure formation process.  In order to understand how
the study of the IGM can help us answer these questions, it is useful
to think of two separate characterizations of the process. First we
can think of the evolution of the average ionization fraction
of hydrogen with time and we can subsequently consider the spatial fluctuations
around that mean evolution. The latter is the study of the sizes of the
ionized bubbles and the morphology of the ionized regions. This split
is useful because different observational probes have different
sensitivities to these different  aspects. Furthermore the study of
the mean and the spatial fluctuations are sensitive to different
properties of the first sources.

\begin{figure}[!ht]
\begin{center}
\mbox{\epsfig{file=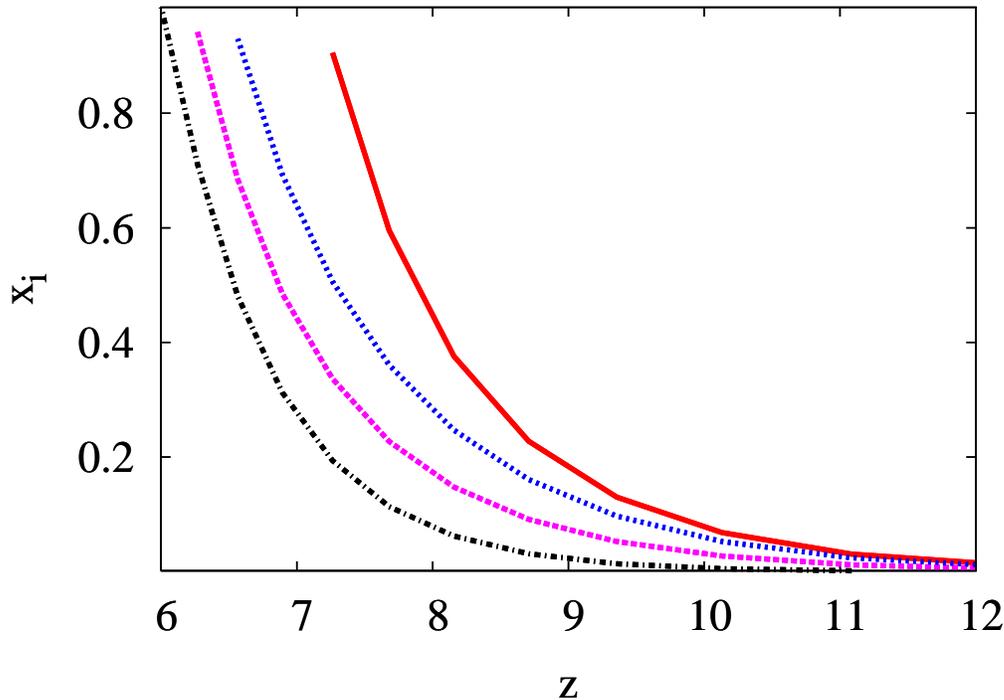,width=10cm,angle=270}}
\caption{Ionization history of the IGM from \cite{McQuinn:2006et}. The curves show example
theoretical models describing the volume-filling fraction of ionized
gas, $x_i$, as a function of redshift. Observational probes of
reionization aim to map-out, or constrain, the detailed dependence of
ionization fraction with redshift. When does essentially the entire
volume of the IGM become filled with ionized gas ($x_i = 1$)? How
extended is the reionization process? Answering these questions will
provide significant constraints on the first generation of ionizing
sources, and on the surrounding IGM. The example curves illustrate the
impact of one relevant effect, radiative, thermal feedback, which acts to
suppress the formation of galaxies within ionized regions. The red
solid curve ignores feedback, the black dashed line incorporates a
`maximal' level of thermal feedback, while the other curves include
intermediate levels of feedback. (See \cite{McQuinn:2006et} for details.) The timing and
duration of reionization depend strongly on feedback processes such
as this.}
\end{center}
\label{fig:xifeedback}
\end{figure}

To learn about what type of halos the sources lived in one can study
how rapid the evolution of the mean ionization fraction is. Sources
that live in rarer more massive halos usually lead to a faster
evolution of the mean. The time and duration of reionization also depends
on the clumpiness of the IGM, as well as on radiative, chemical, and mechanical
feedback from the first sources, which impact the formation of sources
forming later on. An example of the type of information contained in the
redshift evolution of the mean ionization fraction is shown in Figure
1, which demonstrates the impact of radiative feedback.

Furthermore any evidence of ionized hydrogen at very
early times could signal the existence of an early and very efficient
mode of star formation or perhaps an entirely new source of ionizing
photons such as those that could come from annihilating  or decaying
dark matter. The study of large scale CMB polarization is probably the
best and only probe of these very early stages of the ionization
history and thus it is probably the only way to constrain some of
these scenarios.

\begin{figure}[!ht]
\centerline{\epsfxsize=12cm\epsffile{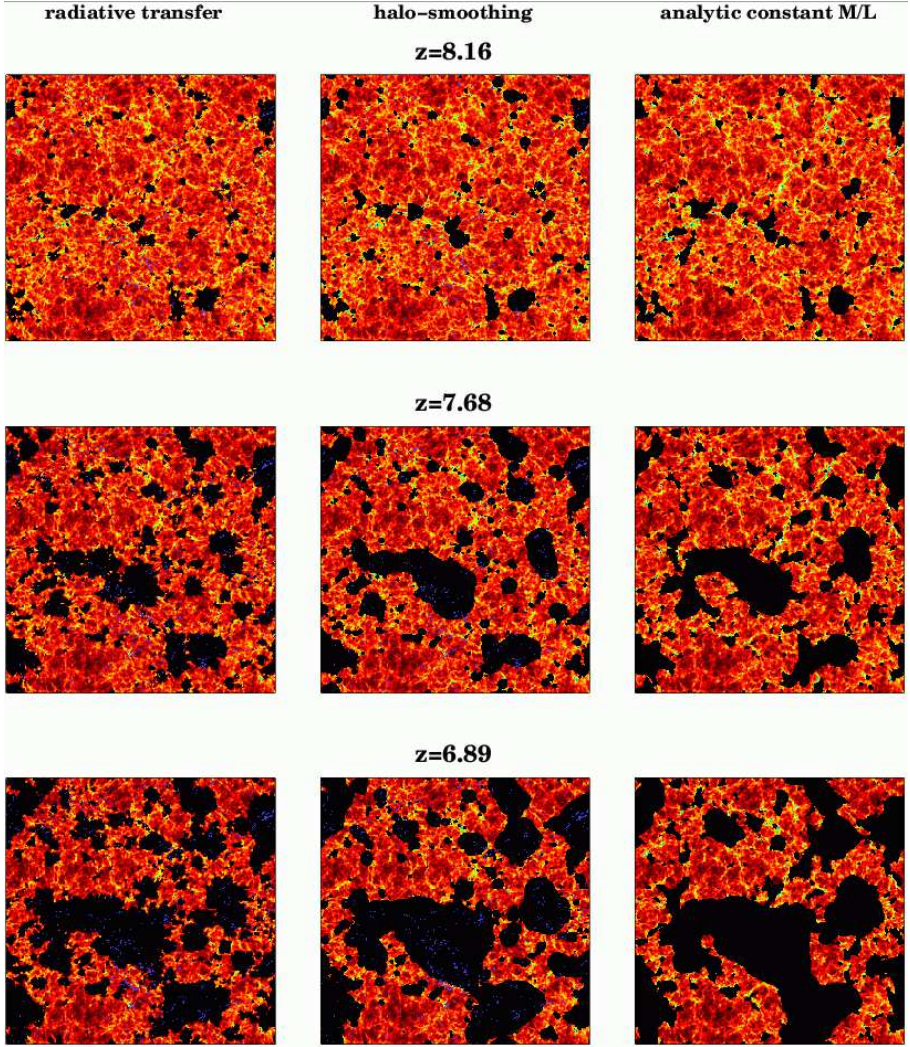}}
\caption{Ionized bubbles during reionization from \cite{Zahn:2006sg}.
 Each panel is a $0.25$
(co-moving) Mpc/$h$ deep slice from a simulation box that is $65.6$
Mpc/$h$ on a side.  The black regions show ionized bubbles and the
color scheme is proportional to neutral hydrogen density.  The panels
show different stages of reionization, with $x_i = 13\%, 35\%$ and
$55\%$ of the volume of the IGM ionized from top to
bottom. Reionization studies aim to map-out, or constrain, the size
distribution of these ionized bubbles at different stages of the
reionization process. The left-most panels show results from a
radiative transfer simulation of reionization, while the center and
right-most panels show results from semi-analytic calculations.  The
blue points in the center and left panels show ionizing sources from
the simulation.}
\label{fig:bubbles}
\end{figure}

The study of the spatial variations of the ionization fraction during
reionization, that is determining the size distribution of the ionized
regions, also encodes a wealth of information about the sources
responsible for reionization. 
A simulated model, illustrating spatial variations in the ionization
fraction, is shown in Figure \ref{fig:bubbles}.
The size distribution is mainly
dependent on the clustering properties of the sources which in turn
can help determine what mass halos those sources typically
inhabit. Furthermore, and especially during the later stages of
reionization, the presence of high density regions that act as photon
sinks can set a maximum size for the ionized bubbles. Thus there is
the potential to learn about the small scale density structure of
the IGM.

\subsection{Outline of this White Paper}

In section \ref{sec:obs_probes} we will describe different probes that can be used to constrain the ionization history of the universe. 
In section \ref{sec:lspol} we focus on the information that can be extracted from the large scale polarization of the CMB. In section \ref{sec:otherparams} we comment on the impact uncertainties in the ionization history have on the determination of other cosmological parameters. Finally for completeness in section \ref{sec:ss_cmb} we discuss the measurement of secondary anisotropies.

\section{Survey of observational probes}
\label{sec:obs_probes}

As mentioned in the introduction, a confluence of data sets from a variety
of different observational probes should soon provide significant advances in our understanding of
the EoR.
In order to best understand how CMBPol may help with
the study of reionization, we briefly describe several future
probes, examining in particular how large scale
polarization measurements may complement constraints from these other
exciting studies. As we will see, upcoming observations should provide
detailed constraints on reionization activity at redshifts close to
and a bit higher than $z \sim 6$ while large scale polarization, and
CMBPol in particular, may be our best way to study reionization at
higher redshifts, $z \gtrsim 10-12$, in the near future.

We describe here four current and upcoming probes of reionization: high redshift quasar
spectra, surveys for Ly-$\alpha$ emitting galaxies, optical afterglow
spectra of gamma-ray bursts, and 21 cm
observations of the high redshift IGM. In \S \ref{sec:ss_cmb} we additionally discuss constraints from small
scale CMB measurements, which will be made by several upcoming experiments, and potentially by CMBPol,
provided its angular resolution is high enough. We also discuss the related topic of metal absorption. 

\subsection{High Redshift Quasar Spectra}
\label{sec:quasars}

Presently, our most detailed probe of the high redshift ($z \gtrsim 6$) IGM
comes from Ly-$\alpha$ forest absorption spectra towards high redshift
quasars. Regions with relatively large $\rm HI$ densities
should appear as absorption troughs in quasar spectra, which presumably
deepen and come to dominate as we approach the reionization epoch.
Indeed, spectra of $z \sim 6$ quasars selected from the SDSS show some extended
regions of zero transmission \cite{becker}. Unfortunately, the Ly-$\alpha$ cross section is very large
and a relatively low average neutral fraction at the level of $\langle X_{\rm HI} \rangle \sim
10^{-3} - 10^{-4}$ is sufficient to give complete absorption in the
$z \sim 6$ Ly-$\alpha$ forest \cite{Fan:2005es}. This makes it difficult to distinguish a
largely neutral IGM from a significantly ionized one with this probe.

Nevertheless, several detailed studies have examined whether these spectra may indeed probe
the IGM before reionization completes. These analyses have considered a range of
different statistics: the redshift
evolution of the average absorption in the Ly-$\alpha$ and Ly-$\beta$ forest, the
amount of scatter in the absorption from sightline to sightline, the abundance and
extent of dark gaps in the high redshift Ly-$\alpha$ forest, the properties of
the proximity zone regions in high redshift quasar spectra, the redshift evolution
of metal abundances, and the temperature of the IGM from the $z \sim 2-4$ Ly-$\alpha$ forest.

\begin{figure}[!ht]
\centerline{\epsfxsize=12cm\epsffile{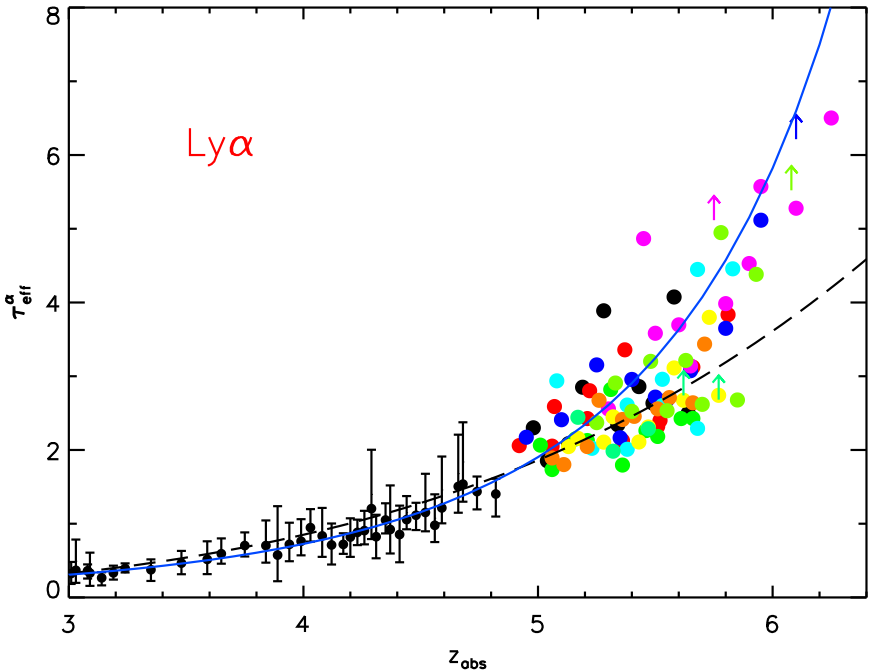}}
\caption{Redshift evolution of the Ly-$\alpha$ forest opacity  from \cite{Becker:2006qj}. The black
points are measurements of the $z \lesssim 5$ Ly-$\alpha$ opacity
from \cite{Songaila04}. The solid colored points are measurements from individual
$\Delta z = 0.15$ stretches of Ly-$\alpha$ forest spectra \cite{Fan:2005es}, with arrows
indicating lower limits on stretches consistent with complete absorption.
The black dashed and solid blue curves indicate different extrapolations
of the $z \lesssim 5.5$ opacity to $z \gtrsim 6$.}
\label{fig:tau_lya}
\end{figure}

A plot of the redshift evolution of the Ly-$\alpha$ opacity is shown
in Figure \ref{fig:tau_lya}. More precisely, we show the redshift 
evolution of the effective Ly-$\alpha$ opacity, defined by
$\tau_{\rm eff} = -\ln \langle F \rangle$, where $\langle F \rangle$ is the
average flux transmitted through the Ly-$\alpha$ forest. One can see that the 
opacity at $z \sim 6$ is typically very large, $\tau_{\rm eff} \gtrsim 5$,
with some sightlines having only lower limits on the opacity -- i.e.,
they are consistent with complete absorption in the $z \sim 6$ Ly-$\alpha$
forest. As mentioned before, complete absorption in the Ly-$\alpha$ forest
at $z \sim 6$ only implies a weak constraint on the average neutral fraction.
A slightly stronger constraint comes from spectra with complete absorption
in the Ly-$\beta$ region of the forest. Since the Ly-$\beta$ cross section is a 
factor of $\sim 6$ times smaller
than that for Ly-$\alpha$ absorption, complete Ly-$\beta$ absorption implies
a more stringent lower limit on the neutral fraction than complete absorption in Ly-$\alpha$, but
the resulting constraints are still consistent with a mostly ionized IGM \cite{Fan:2005es}.

A second feature of this figure, however, is that the opacity appears to
evolve rather sharply with redshift around $z \sim 6$. Perhaps the sharp evolution in the
opacity is signaling correspondingly rapid changes in the ionization state
of the IGM?  This is currently a matter of debate. In \cite{Fan:2005es} the authors  fit a power-law
to $z \lesssim 5.5$ data (shown as the black dashed line in Figure \ref{fig:tau_lya}), and 
find that the $z \geq 5.7$ data show a rapid upward departure
from this fit \cite{Fan:2005es}. Using a simulated model for the IGM density distribution \cite{Miralda00}, they argue that
this upward departure requires very rapid evolution in the ionization state of the IGM. On the
other hand, the authors of \cite{Becker:2006qj} use an empirically motivated fit to the opacity distribution 
and claim that rapid changes in the ionization state are not required \cite{Becker:2006qj}. The blue solid line in
Figure \ref{fig:tau_lya} shows an extrapolation of their model from $z \leq 5.4$ data to higher
redshift, assuming a slowly varying ionization state. This model also fits the 
$z \gtrsim 6$ Ly-$\alpha$ opacity data.
The main issue here is that, even in a highly
ionized IGM, the Ly-$\alpha$ absorption cross section is so large that only quite underdense regions
allow transmission through the $z \sim 6$ Ly-$\alpha$ forest. This means that the opacity
evolution is sensitive to the precise abundance of underdense regions, as are the constraints
on the ionization state (\cite{Oh05}, \cite{Lidz06a}). Further theoretical modeling may help resolve this debate.

In addition to the large average absorption, the scatter in the absorption from sightline to sightline
is quite large at $z \sim 6$ \cite{Fan:2005es}. Indeed, the colored points in Figure \ref{fig:tau_lya} show order unity
variations in the opacity, after averaging over $\Delta z = 0.15$ or $\sim 50$ co-moving Mpc/$h$. 
These variations might indicate incomplete and patchy reionization with some sightlines piercing through
mostly ionized gas, and others passing through long neutral stretches \cite{Wyithe:2005sy}. However, variations in the line
of sight density field give rise to surprisingly large opacity fluctuations at $z \sim 6$,
and so it is difficult to discern, on the basis of this scatter, whether these spectra really probe the IGM 
before reionization completes \cite{Lidz06a}. Current measurements appear at least broadly consistent with density fluctuations alone \cite{Lidz06a, Liu06}.

The tightest constraints claimed on the ionization state of
the $z \sim 6$ IGM come from measurements of the proximity regions
around $z \sim 6$ quasars.  Several authors, starting with \cite{Wyithe:2004mx}, 
have argued that these regions are small,
indicating that quasar ionization fronts are expanding into a largely
neutral IGM \cite{Wyithe:2004mx, Wyithe:2004jw, Mesinger:2006kn}. However, 
subsequent more detailed studies have shown that
it is hard to distinguish the case of a quasar ionization front expanding
into a partly neutral IGM, and the case of a mere proximity zone in a mostly
ionized IGM (e.g., \cite{Bolton:2006pc, Lidz:2007mz}). Detailed modeling of inhomogeneous reionization, incorporating
the tendency for quasars to live in overdense regions that are ionized before
typical regions, is required to best interpret these observations \cite{Lidz:2007mz}.

One can also measure the redshift evolution of metal line abundances in high redshift quasar spectra,
and use this to constrain the reionization history.
The mass density in Carbon IV appears to be a surprisingly flat function of redshift from 
$z \sim 2-6$ \cite{Ryan-Weber:2006, Simcoe:2006}, requiring significant levels of star 
formation and metal enrichment prior to $z = 6$. Another metal line tracer is the OI line 
which has an ionization potential similar to that of hydrogen, lies redward of Ly-$\alpha$, and has
a significantly lower optical depth than Ly-$\alpha$, making it a potential probe of the IGM
before reionization completes \cite{Oh:2002}. In fact, high-resolution Keck spectra of the
$z \sim 6$ SDSS quasars do reveal some OI lines, with $4$ out of $6$ detected systems
lying towards the highest redshift quasar presently known \cite{Becker:2006b}. Interestingly, some
of the OI systems are nearby regions that show transmission in the Ly-$\alpha$ and Ly-$\beta$ forests
of this quasar \cite{Becker:2006b}.
The interpretation of these observations is
unclear: the OI systems might reflect dense clumps of neutral gas in a
highly ionized IGM, or instead could indicate inhomogeneous metal
pollution in a more neutral IGM.

A final constraint comes from measurements of the temperature of the
Lyman-$\alpha$ forest at $z \sim 2$--$4$, which suggest an order unity
change in the ionized fraction at $z_r <10$
\cite{theuns02-reion,hui03}, although this argument depends on the
timing and history of $\rm HeII$ reionization (e.g.,
\cite{sokasian02, McQuinn:2008am}).

Some progress will be made in the near future by detecting more quasars,
some at higher redshifts ($z \gtrsim 6.5$), using widefield, deep near-IR surveys, although
the rarity of bright background quasars at these redshifts makes this
challenging. Improvements in modeling inhomogeneous reionization should also help our understanding of
$z \gtrsim 6$ absorption spectra \cite{Kohler:2005gg, Lidz:2007mz, Trac:2008yz}.
However, the large Ly-$\alpha$ opacity expected in a mostly ionized medium is clearly a fundamental limitation, and
it is likely impossible to study all but the final stages of reionization with this approach.

\subsection{Lyman-alpha Emitters}

An effective way to find high redshift galaxies is to search for their Ly-$\alpha$
emission lines, since young galaxies frequently have strong Ly-$\alpha$ emission \cite{Partridge67}. Ly-$\alpha$ emission 
searches have an advantage over high redshift
Lyman break surveys in that they target narrow wavelength intervals,
in between strong night sky background lines, in search of prominent emission
lines. This allows one to detect galaxies that are unobservable by Lyman
break selection owing to the strong night sky background at relevant
wavelengths.
Existing surveys have discovered several hundred galaxies with
this technique (e.g., \cite{hue02,kodaira03,
rhoads04, stanway04,santos04-obs, kashikawa06, cuby06, willis05,
stark07a}), and other programs will begin taking data soon (e.g.,
\cite{casali06, mcpherson06, horton04}). The largest existing high redshift
sample from the Subaru Deep Field consists of approximately $50$ 
galaxies above $z \sim 6.5$ \cite{kashikawa06}.

\begin{figure}[!ht]
\centerline{\epsfxsize=12cm\epsffile{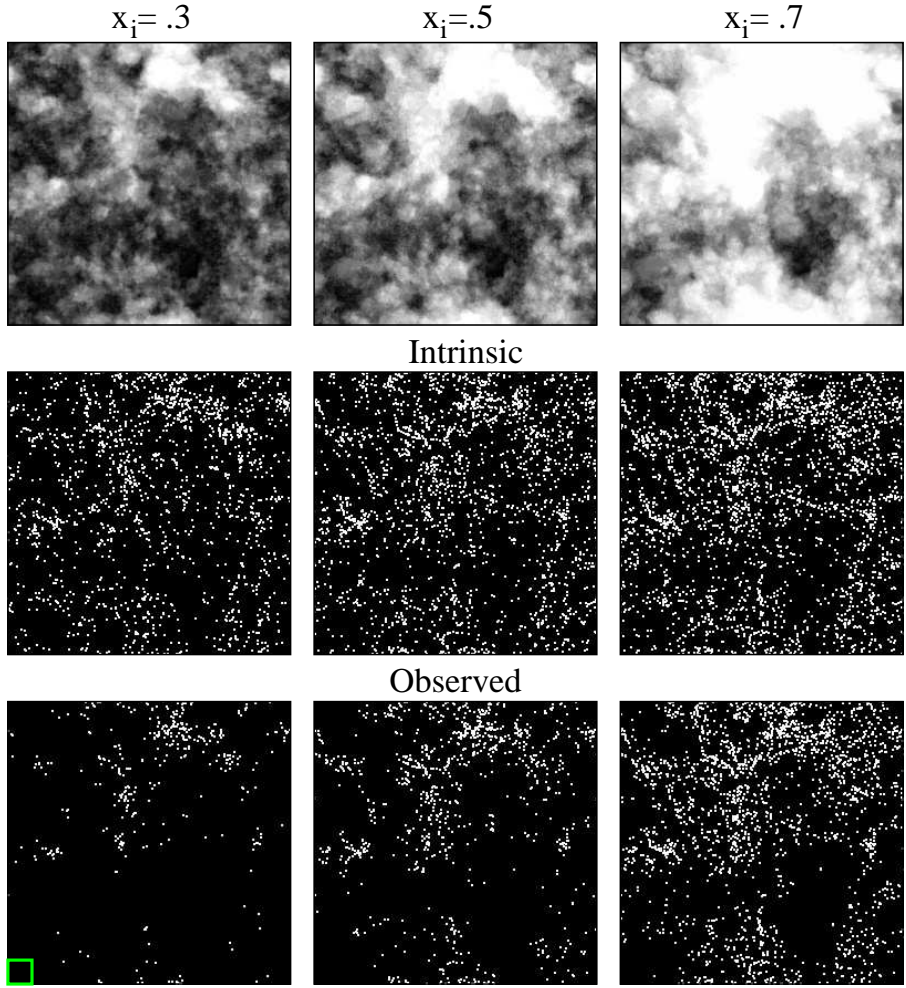}}
\caption{Ly-$\alpha$ emitting galaxies during the EoR. The top panel
shows maps of the ionization field from a simulation at three
different (volume-averaged) ionized fractions: $\langle{x_i}\rangle=0.3, 0.5,$ and
$0.7$ from left to right respectively. The middle panel shows the
true distribution of Ly-$\alpha$ emitting galaxies from the corresponding
simulation slices. The bottom panel shows the {\em apparent} distribution
of the same galaxies, based on observing the sources' Ly-$\alpha$ emission
lines. From \cite{McQuinn07b}.}
\label{fig:lae}
\end{figure}

During the EoR, neutral hydrogen gas in the IGM will extinguish galactic Ly-$\alpha$ emission lines.
Specifically, the optical depth to Ly-$\alpha$ absorption in a significantly neutral IGM is so large 
that even the red side of the Ly-$\alpha$ line is 
attenuated by damping wing absorption \cite{Miralda98}. As a result, galaxies that 
lie sufficiently close to the edge of an HII region, next to a significant
column of neutral hydrogen, will be unobservable in Ly-$\alpha$. On the other
hand, galaxies which reside towards the center of a sufficiently large
HII region avoid complete attenuation, and will still be visible in Ly-$\alpha$
emission. Since large HII regions form around high density peaks, one
expects the clustering of Ly-$\alpha$ emitters to increase dramatically
towards the start of reionization \cite{Furlanetto06c,McQuinn07b,Mesinger07b}. In conjunction, the 
abundance of Ly-$\alpha$ emitters should drop off rapidly as one looks back deeper into the EoR. 

These trends can be seen in the simulation
slices of Figure \ref{fig:lae}. Comparing the top, middle, and bottom panels, one can see how the presence
of ionized bubbles modulates the abundance of emitters observable in Ly-$\alpha$: sources close to the
center of large ionized regions are observable, while other sources 
are attenuated out of the Ly-$\alpha$ sample. Comparing the left-most and right-most panel,
it is clear that the suppression is much more significant in the earliest stages (here when the volume ionized
fraction is $x_i = 0.3$), compared to later stages (here when $x_i = 0.7$). It is also clear that -- since
the ionized regions quickly grow rather large -- one needs to observe a large field of view to reliably
measure the abundance and clustering of Ly-$\alpha$ emitters during reionization. To best constrain
reionization with this technique, one hence wants a widefield survey that is deep enough to find large samples
of emitters in the early stages of reionization.

Several authors have, in fact, used existing Ly-$\alpha$ emitter samples to place constraints
on the ionization state of the IGM. Malhotra \& Rhoads \cite{Malhotra06} constrained the ionized volume
fraction in the IGM by counting the abundance of Ly-$\alpha$ emitters at $z=6.6$, and requiring
a minimum ionized volume around the sources to avoid attenuating their Ly-$\alpha$ photons. 
From this argument, they find that more 
than $50\%$ of the volume of the IGM is ionized at $z = 6.6$.
Kashikawa et al.  \cite{kashikawa06} observe a $\sim 2-\sigma$ decline in the bright end of the luminosity
function between $z=5.7$ and $z=6.5$, which they attribute to incomplete reionization at
$z \gtrsim 6$. Several other authors argue, however, that the observed 
evolution is consistent with evolution in a 
post-reionization IGM \cite{McQuinn:2006et,Dijkstra07,Fernandez08}. Finally, McQuinn et al. \cite{McQuinn07b}
use the clustering of the $z=6.6$ Subaru Deep Field sample to exclude a significantly neutral Universe
with $x_i \leq 0.5$ at $2-\sigma$ confidence. 

Further progress can be expected, with numerous planned surveys attempting to find 
Ly-$\alpha$ emitters at high redshift. In particular, an extension to the existing 
Subaru Deep Field survey should
extend measurements to $z = 7.3$ over a $\sim 2$ deg$^2$ field of view. This amounts to 
a factor of $\sim 10$ boost
in field of view compared to the existing Subaru Deep Field. It will likely be difficult to
push this method to significantly higher redshifts, since strong night sky emission lines force one to make 
near IR measurements from space, and since we require observations over a wide field of view. We expect
this method to be most interesting in the near future if a significant fraction of 
the volume of the IGM is still
neutral near $z = 7$.


\subsection{GRB Optical Afterglows}

Another possibility is to search for damping wing absorption in high
redshift GRB optical afterglow spectra \cite{Miralda98,Barkana04,Totani:2006,McQuinn:2007gm}. These sources are extremely
luminous, can potentially be detected at very high redshift, and have
the important advantage of a simple power-law intrinsic spectrum. Additionally,
GRBs are more likely to exist in typical HII regions than quasars or the
massive, highly luminous galaxies that are easiest to detect at high redshift. 
Thus it may be possible to identify damping wing absorption in individual
GRB afterglow spectra. 

\begin{figure}[!ht]
\centerline{\epsfxsize=12cm\epsffile{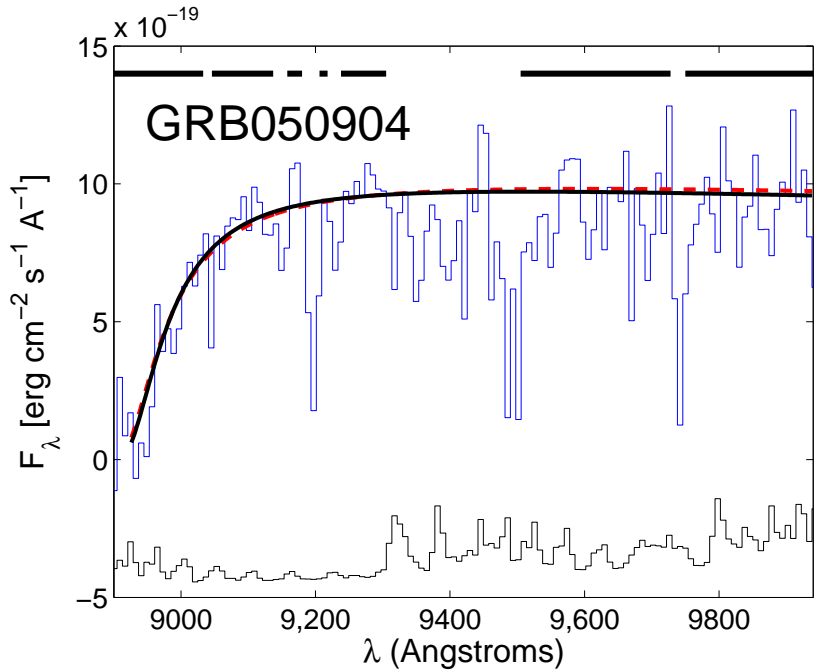}}
\caption{GRB optical afterglow spectra and DLAs. The blue line shows a portion
of the observed spectrum of the GRB optical afterglow spectrum of GRB050904,
at $z=6.3$ (from \cite{Totani:2006}). The black solid line is a model spectrum, consisting entirely
of absorption from a DLA associated with the GRB host galaxy, of column
density $N_{\rm HI} = 10^{21.6} cm^{-2}$. The red dashed line is a model
that includes both DLA absorption and damping wing absorption from a 
completely neutral IGM.
The presence of a large column density local absorber makes
it extremely difficult to tell whether there is absorption from neutral 
hydrogen in the surrounding IGM. From \cite{McQuinn:2007gm}.}
\label{fig:grb_dla}
\end{figure}

A difficulty with this probe of reionization, however, is that many optical
afterglows show damped Ly-$\alpha$ absorption from neutral hydrogen in
the GRB host galaxy, at a level that is large enough to overwhelm the damping wing
opacity from neutral gas in the surrounding IGM. For example, the
highest redshift GRB afterglow detected thus far, at $z=6.3$ \cite{Totani:2006}, shows evidence for
a strong Damped Ly-$\alpha$ Absorber (DLA) in its host, with a column density of 
$N_{\rm HI} = 10^{21.6} {\rm cm^{-2}}$. This absorption (shown in Figure 
\ref{fig:grb_dla}) is so strong that it is difficult to tell whether or not
there is damping wing absorption from the surrounding IGM.
The damping wing profile from extended neutral gas in the IGM has a different wavelength
dependence than that from a compact, local absorber, and we can therefore still hope to unambiguously
detect neutral gas in the IGM using GRB afterglow spectra \cite{Miralda98}. To detect the IGM damping wing, a GRB
with a weaker local absorber, smaller than roughly $N_{\rm HI} \lesssim 10^{20} cm^{-2}$, is needed.
In addition, a high signal-to-noise spectrum is desirable, demanding rapid near IR follow-up
of probable high redshift bursts. Some afterglow spectra at lower redshift show local
absorbers with significantly
smaller column density than the $z=6.3$ burst, and occasional spectra show no
DLAs whatsoever \cite{Chen:2007}. We may therefore hope to find a clean line of sight towards a GRB 
at higher redshift and apply this test.

\begin{figure}
\begin{center}
\mbox{\epsfig{file=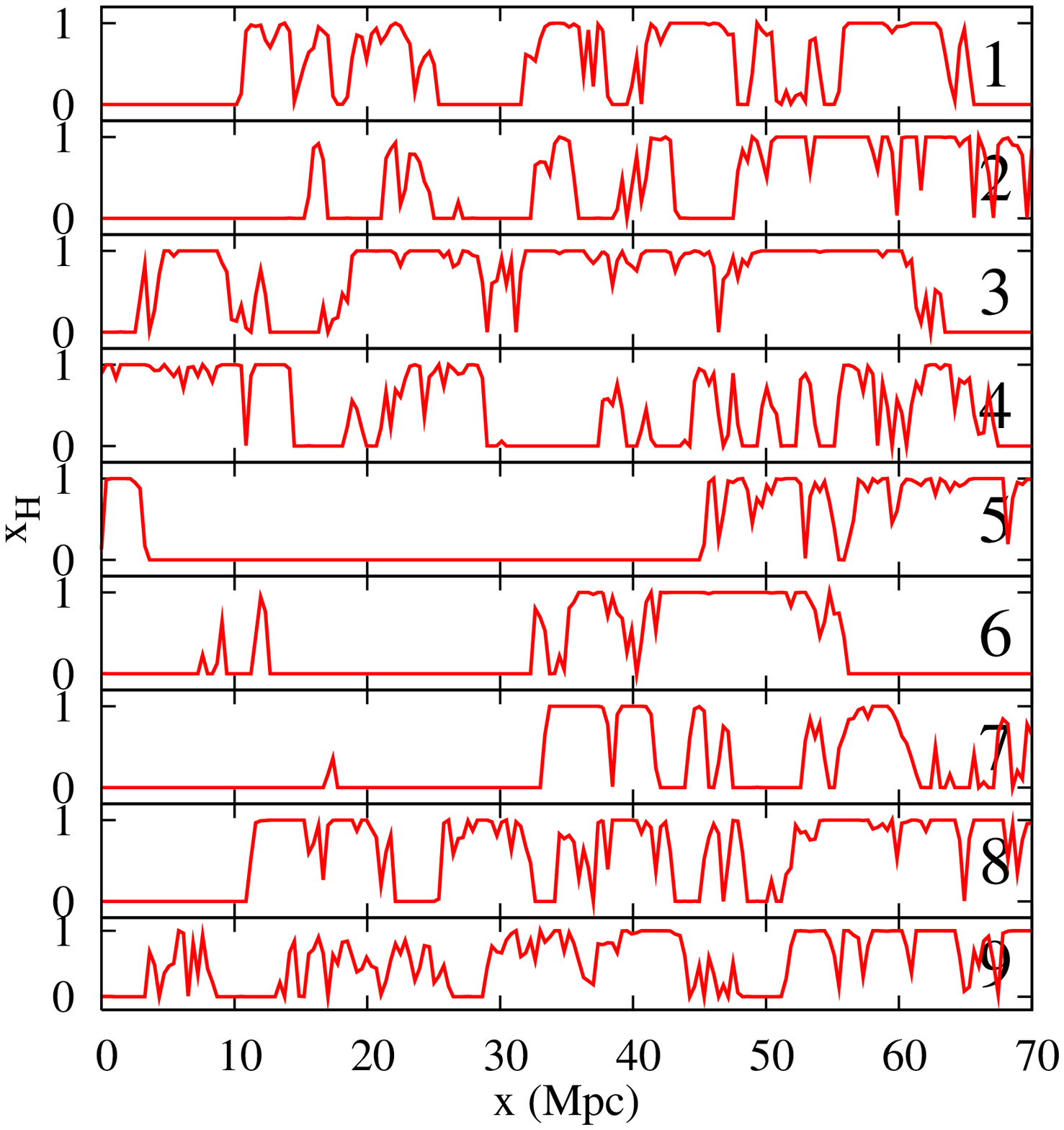,width=8.5cm,angle=270}}
\mbox{\epsfig{file=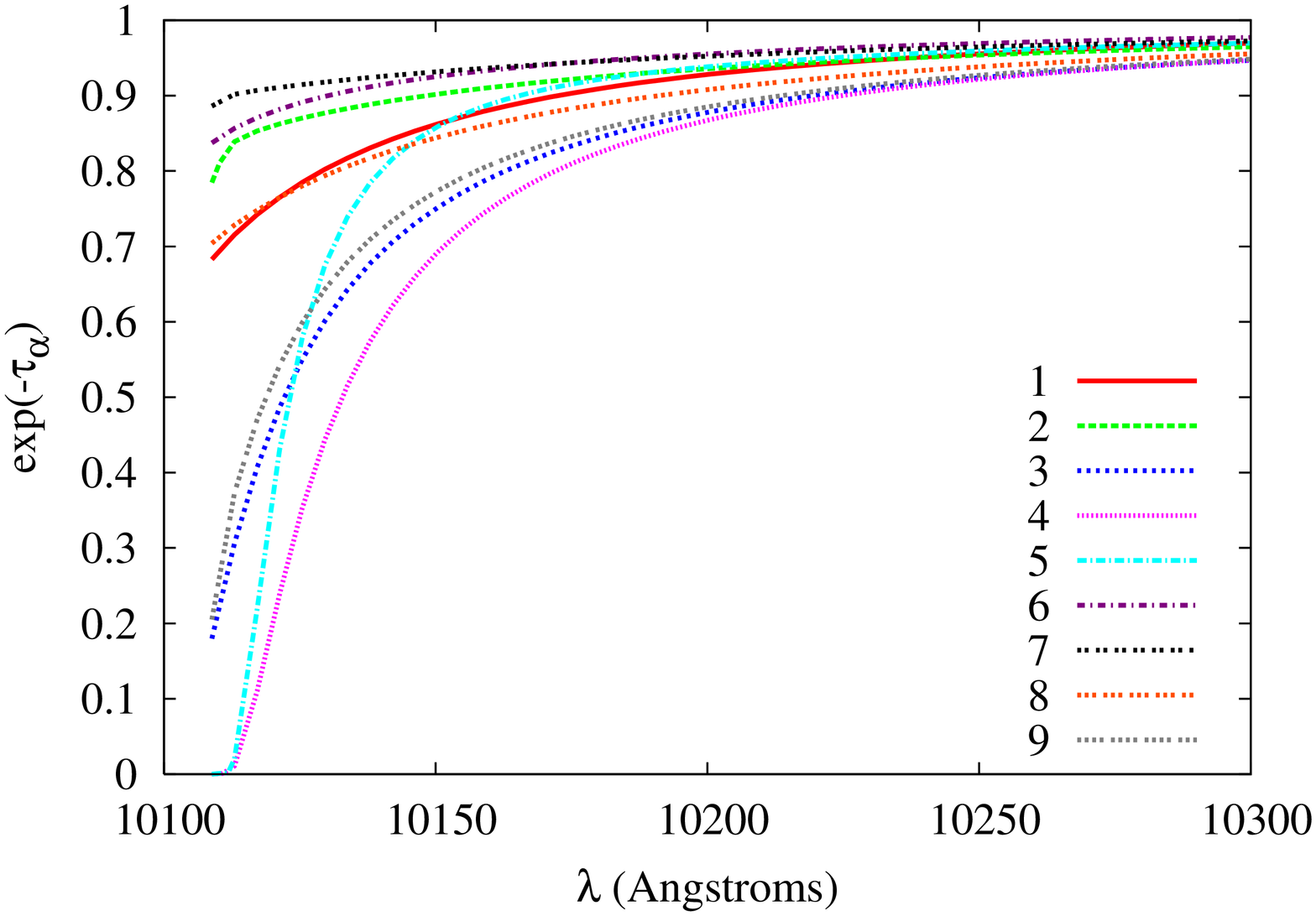,width=8.5cm,angle=270}}
\caption{Sample variance and GRB afterglow spectra from \cite{McQuinn:2007gm}. {\em Top panel}: The lines
show the neutral fraction along several lines of sight towards a GRB host, when
the IGM is $\sim 50\%$ ionized by volume. {\em Bottom panel}: The 
damping wing absorption vs. wavelength for each of the nine sightlines
shown in the top panel.
Some lines of sight lead
to significant levels of damping wing absorption, while other lines of
sight result in very little damping wing absorption. It is hence difficult
to discern the ionized fraction from a few optical afterglow spectra, even
if they unambiguously show damping wing absorption from the IGM}
\label{fig:grbvar}
\end{center}
\end{figure}

While GRB optical afterglow spectra may provide a good way of detecting 
the {\em presence} of neutral hydrogen in the IGM, perhaps even at very
high redshift, it will be difficult to extract the ionized fraction from
such measurements. The challenge here is that reionization is expected
to be extremely {\em inhomogeneous} and the level of damping wing absorption
should vary considerably towards one GRB host or another. Some GRBs will reside
in large HII regions even rather early in the reionization process, since large
ionized regions can form under the collective influence of highly clustered
neighboring galaxies. The GRBs in large HII regions will suffer little damping wing
absorption, while GRBs in smaller ionized bubbles or close to the edge of a large
bubble will suffer more damping wing absorption \cite{McQuinn:2007gm}. 
This is illustrated
in Figure \ref{fig:grbvar}, which shows that even when the IGM is as
much as half neutral, there will be significant scatter in the level of
damping wing absorption. This is similar to the effect 
depicted in Figure \ref{fig:lae},
but here we are attempting to extract information based on the spectrum of 
an individual object, rather than statistical trends based on the abundance and clustering of 
an entire population of galaxies. Extracting the ionized fraction from this method may
require a prohibitively large sample of high redshift afterglows. We are hopeful
that this method will definitively detect neutral gas at high redshift, indicating that
reionization is not yet complete at a given redshift, but expect other probes to best
map out the redshift evolution of the ionized fraction.




\subsection{The 21 cm Line}

One of the best ways to study reionization may be to detect redshifted 21 cm emission from neutral
hydrogen gas in the high redshift IGM. Whenever the spin
temperature of neutral hydrogen is different from the CMB temperature,
neutral hydrogen atoms can be seen either in emission or absorption
against the CMB at the redshifted wavelength of their 21 cm
transition. After the first sources appear, the gas is expected to
have been heated and the kinetic temperature of the gas should exceed
that of the CMB \cite{chen03}. Furthermore, Ly-$\alpha$ photons from the first sources
should couple the spin temperature of the 21 cm transition to the gas temperature
long before the IGM is significantly ionized \cite{ciardi03}.  Thus, hydrogen should appear in
emission during the EoR and high resolution observations of the 21 cm
transition as a function of both frequency and angle can provide a
three-dimensional map of reionization (e.g., \cite{Furlanetto:2004nh,Iliev08}).  Several
experiments are now underway or being planned to measure this signal,
including the 21 cm ARRAY to be operated in China (formerly known as
PAST\footnote{http://web.phys.cmu.edu/$\sim$past/}), the Low Frequency
Array (LOFAR\footnote{http://www.lofar.org}), the Murchison Widefield
Array (MWA\footnote{ http://web.haystack.mit.edu/MWA/MWA.html}), the
PAPER\footnote{http://astro.berkeley.edu/~dbacker/EoR/} experiment, an
effort at the GMRT\footnote{Ue-Li Pen private communication.}
telescope, and ultimately the Square Kilometer Array (SKA\footnote{
http://www.skatelescope.org for details on the SKA}).
This method has several key advantages: the optical depth
is low and so one does not suffer from the saturation problems that plague Ly-$\alpha$ forest studies, 
it does not require a high redshift background source, and the probe is a spectral line, allowing
one to extract full three-dimensional information about reionization. For a review, see
\cite{Furlanetto06a}. 

Futuristic experiments like the SKA should have the sensitivity
to produce relatively detailed maps of the reionization process. These
maps, taken at several different frequencies, will amount to a reionization
`movie': they will depict the growth of HII regions around individual sources,
their subsequent mergers with neighboring HII regions, and 
detail the completion of the reionization process, whereby
the entire volume of the Universe becomes filled with ionized gas.
Reionization maps from the SKA should rather directly reveal the
size distribution of ionized regions during reionization, and the
redshift evolution of the ionized fraction. 
First generation 21 cm surveys,
such as the MWA, LOFAR, GMRT, and the 21 cm ARRAY, will not have the sensitivity
to make detailed maps of the reionization process, but will allow for
a statistical detection \cite{McQuinn:2005hk}. For example, the MWA should measure the power
spectrum of 21 cm fluctuations over roughly a decade in scale, in each
of several independent redshift bins \cite{Lidz:2007az}. These measurements will already
be quite valuable, but may not be entirely straightforward to interpret.

\begin{figure}[!ht]
\centerline{\epsfxsize=12cm\epsffile{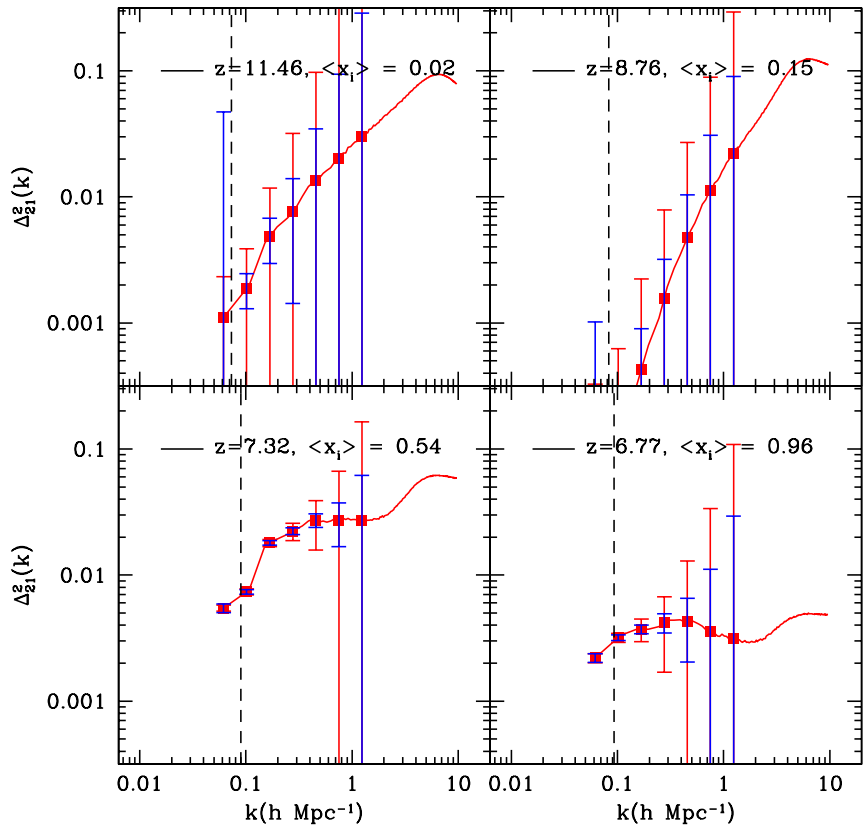}}
\caption{Sensitivity of the MWA for measuring the 21 cm power spectrum.
The red curves show dimensionless 21 cm power spectra from a reionization model
at different redshifts and ionization fractions, as labeled. The
red error bars indicate statistical error estimates for $1,000$ hours
of observations. The blue error bars indicate error estimates for a different
configuration of the MWA antennas. Foreground cleaning will likely
prohibit power spectrum measurements at wavenumbers smaller than
indicated by the black dashed lines. From \cite{Lidz:2007az}.
}
\label{fig:21cm_power}
\end{figure}

Here we briefly discuss what we might hope to learn from these first generation experiments, which
will come online before CMBPol.
A model for the 21 cm power spectrum during reionization is 
shown in Figure \ref{fig:21cm_power}, along with approximate
error bars for one year of MWA observations \cite{Lidz:2007az}. At early times -- as seen in the top left hand panel
of the figure -- before significant ionized regions grow in the model, 
the 21 cm power spectrum traces the density power spectrum.\footnote{Spin temperature fluctuations are
neglected in this model. Including these will modify the predictions in the early stages of 
reionization 
(e.g., \cite{Pritchard07}).}
As large ionized bubbles grow around early generations of galaxies, the amplitude of the 21 cm
power spectrum increases on large scales, and the slope of the 21 cm power spectrum on MWA
scales flattens. This behavior is seen in the bottom left hand panel of Figure \ref{fig:21cm_power}. 
Finally, the amplitude of 21 cm fluctuations drops as the Universe becomes still more ionized
and neutral hydrogen becomes scarce (bottom right hand panel). The amplitude on MWA scales
is typically maximal around the time when the Universe is half ionized. To recap, one expects
the amplitude of the power spectrum to rise and fall as the Universe becomes ionized, and the slope
of the power spectrum to flatten as ionized regions become large and impact the scales probed 
by the MWA. 
These trends should be robust, but detailed modeling is clearly required to best extract information
about the redshift evolution of the ionized fraction and the size distribution of ionized regions
during reionization. Note that the 21 cm signal should be highly non-Gaussian during the EoR:
there is more information than contained in the power spectrum alone and there
may be better ways to extract constraints on the ionized fraction and bubble sizes from
the observations.

The figure also shows error bar estimates for about one year of
observations with the MWA at different redshifts. On small scales, the
detector noise becomes large because the instrument has few antennas
at long baselines and it's high $k$ sampling is thus  mainly limited
to modes in the frequency direction. Large scale modes, on the other
hand, will be  lost to foreground cleaning.  Astrophysical foregrounds
are a considerable challenge for these observations, as the
foregrounds are expected to be four orders of magnitude larger than
the signal itself. Fortunately, known astrophysical foregrounds are
spectrally smooth and hence distinguishable from the high redshift 21
cm signal itself (e.g., \cite{Zaldarriaga:2003du}). Nonetheless, one inevitably loses long wavelength
modes along the line of sight in the foreground cleaning process. A
rough estimate is that foregrounds preclude measuring 21 cm power at
wavenumbers short of the dashed line shown in the figure. This implies
that the MWA is sensitive to roughly a decade in scale, probing
wavenumbers  between $k \sim 0.1 -1 h$ Mpc$^{-1}$. The red and blue
points show error estimates for different configurations of the MWA's
antennas (see \cite{Lidz:2007az} for details).

Another
important point for our present discussion is that the sensitivity is a very strong function 
of redshift,
likely prohibiting upcoming measurements from probing very
high redshifts. The reason for this is that the radio sky is much
brighter at low frequency, with the sky brightness scaling as $T_{\rm
sky} \propto (1+z)^{2.6}$, which results in a rapid increase in
detector noise towards low frequencies, and in limited high redshift
sensitivity. This trend is evident in Figure \ref{fig:21cm_power}. 
Part of the reduced sensitivity towards high redshift results because reionization
occurs relatively late in this model, and the signal is weak at high redshift here. The signal
is typically maximal in the middle of reionization, and the high redshift detectability 
will be boosted 
in models where the middle of reionization occurs at higher redshifts than considered 
here. The increased noise
at high redshift is unavoidable, however, and upcoming surveys will not be sensitive to very high
redshift stages of reionization, losing sensitivity at $z \gtrsim 10-12$. Moreover, because of the 
loss of sensitivity towards high redshift,
first generation surveys will likely focus their efforts exclusively on relatively low redshifts. 
The enormous data rates involved with correlating large numbers of antenna tiles will limit the 
observing bandwidth of the first generation experiments.
The MWA, for example, will process an instantaneous bandwidth of $32$ Mhz, corresponding to a redshift
extent of $\Delta z = 1.8$ near $z \sim 8$. The full spectral range covered by the instrument is much larger,
$80-300$ Mhz, but one does not observe the full spectral range
for free. This fact, and the reduced sensitivity towards high redshift, mean that initial MWA 
observations will likely focus on contiguous $32$ Mhz stretches near $z \gtrsim 6$.

To summarize, we expect exciting constraints on reionization from several independent probes in the next 
few years. These
probes should help determine the redshift evolution of the ionized fraction, and the size distribution of ionized
regions during reionization, which will in turn constrain models for the ionizing sources and early structure
formation. A common feature of each of these probes is that constraining reionization activity at redshifts
close to $z \sim 6$ is much easier then probing reionization at higher redshifts, $z \gtrsim 10-12$.

\section{Large Scale CMB Polarization}
\label{sec:lspol}


The latest WMAP measurements of the large scale CMB polarization constrain the redshift of reionization to be  $z_{\rm reion}=11.0 \pm 1.4$  \cite{Dunetal08}, assuming instantaneous reionization. As large scale polarization measurements improve, this assumption may be relaxed and further details of the ionization history will be constrained by the data. We explore how this may come about in the current section. We start by discussing constraints on the total optical depth and progressively include more parameters to assess the amount of information that a future mission such as CMBPol could obtain.


\subsection{Future Constraints on the optical depth}
\label{section:tau_constraints}

Given the improved sensitivity, Planck is expected to be cosmic
variance limited in polarization out to $l \simeq 10$, improving on
the determination of $\tau$ from WMAP5 by a factor 2.5: $\sigma_\tau =
4.7 \times 10^{-3}$ \cite{ColPierPri08}.  This improvement does not
significantly depend on whether a general reionization history or an
instantaneous reionization scenario is assumed \cite{ColPie08,MorHu08c}.

Planck results may be improved with the help of sub--orbital
experiments observing a large fraction of the sky with low
sensitivity. Indeed, combining Planck with a cosmic--variance limited
experiment out to $l=20-30-50$ and covering half of the sky would
further reduce the error on $\tau$ to $\sigma_\tau = 4.2 - 4.0 - 4.0
\times 10^{-3}$. In this case, sky coverage rather than sensitivity is
the main limitation. More realistic calculations for planned
experiments suggest that these expectations are not affected by actual
noise properties or scanning strategies \cite{Crill08}.

Assuming an instantaneous reionization process a full--sky cosmic
variance limited experiment to $l=2500$ on would have
$\sigma_\tau = 1.9 \times 10^{-3}$: more than a factor two lower than what
Planck will obtain. 
Non--negligible noise and beam effects can degrade
this performance; results for proposed configurations (\cite{EPICpap})
are presented in table \ref{tab:epictau}.

\begin{figure*}
\centerline{
\includegraphics[width=10.cm]{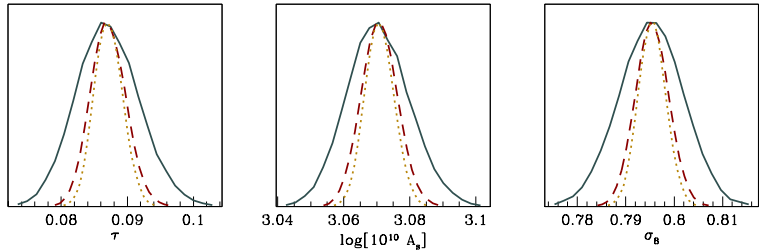}
}
\caption{Marginalized distributions for $\tau$ and related parameters,
for a fiducial model with instantaneous reionization and $\tau =
0.087$ from \cite{EPICpap}.  Curves show the expected performance of Planck (green/solid
lines), CMBPol (the EPIC 2m configuration, red/dashed lines) and a
full sky cosmic variance limited experiment (yellow/dotted lines). For
both Planck and CMBPol we assumed a 80\% sky coverage. Curves include the 
effect of marginalization over the minimal $\Lambda$CDM
parameter set. 
}
\label{fig:sharp_1D}
\end{figure*}

\begin{table}
\centerline{
\begin{tabular}{lcccc}
\hline
                    & Planck & CMBPol & Cosmic Variance \\
\hline
$\tau    $          &$4.7 \times 10^{-3}$    & $2.5 \times 10^{-3}$ & 
$1.9 \times 10^{-3}$   \\
${\cal A}_s $& $9.4 \times 10^{-3}$& $5.1 \times 10^{-3}$ & 
$4.3 \times 10^{-3}$    \\
$\sigma_8$          & $6.7 \times 10^{-3}$   & $3.3 \times 10^{-3}$ &
$2.6 \times 10^{-3}$    \\
\hline
\end{tabular}
}
\caption{MCMC error estimates for $\tau$ and related parameters,
assuming instantaneous reionization \cite{EPICpap}}
\label{tab:epictau}
\end{table}

\subsubsection{Constraints on the timing of reionization}
\label{sec:tau_implications}

\begin{figure*}
\centerline{
\includegraphics[width=12.cm]{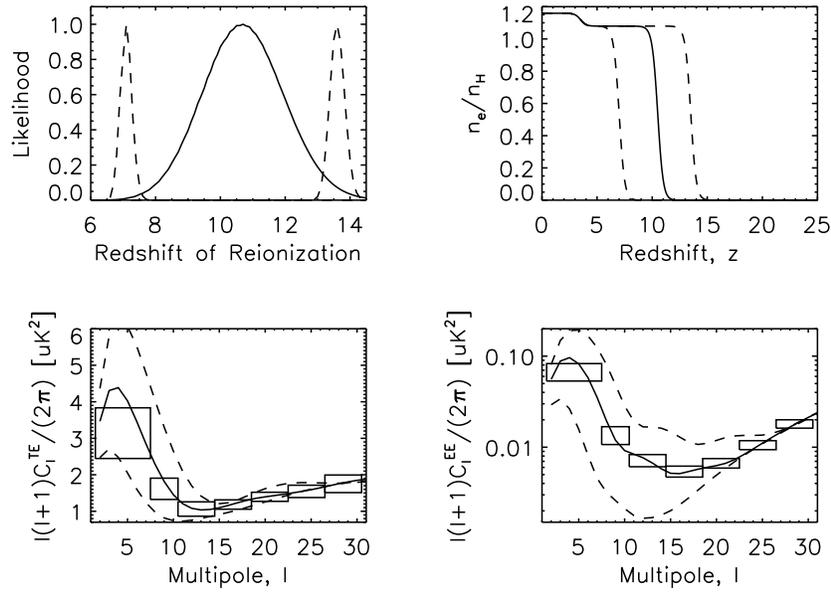}
}
\caption{ Improved constraints on reionization history with CMBPol. {\em Top left}: The solid line shows constraints
on the reionization redshift from WMAP data, after marginalizing over the duration of reionization. The dashed lines are 
estimates for the constraints obtainable with CMBPol in each of a high
redshift and a low redshift reionization model. {\em Top right}: The ionization history in each fiducial model. 
{\em Bottom left}: The $TE$ cross power
spectrum in each fiducial model, along with CMBPol error estimates. {\em Bottom Right}: The same for the $EE$ power spectrum.
}
\label{fig:tau_reion_const}
\end{figure*}

The tight error bar on $\tau$ expected for CMBPol should translate into a strong constraint
on when reionization happens, constraining models for the ionization history and early structure formation, such as those of Figure 1. 
In order to illustrate this, we consider a simple two parameter model 
for the ionization history: the redshift of reionization, $z_r$,
is the redshift at which the hydrogen is half neutral, and ``the
duration of reionization,'' $\Delta$, which is the width of a tanh
function that describes its evolution:
\begin{equation}
 x_e(z) \equiv \frac{n_e(z)}{n_H(z)} =
  \frac{f}2\left\{1+\tanh\left[\frac{(1+z_r)^{3/2}-(1+z)^{3/2}}{\Delta}\right]
	   \right\},
\label{eq:tanh}
\end{equation}
where $f=1+n_{H_e}/n_H\sim 1.08$. This is currently the default
parametrization of {\sf CAMB} code. 
Note that there is no real physical significance to this parametrization, it is merely mathematically convenient. 
See Eq.(95) on pp.14 of CAMB
notes\footnote{{\sf http://cosmologist.info/notes/CAMB.pdf}} for more
details. Note that {\sf CAMB} adds electrons from ionization of HeII,
which increases $x_e(z)$ by about 10\% at $z\lesssim 3.5$. This has
a negligible contribution to the optical depth, however,
since most of the contribution to $\tau_{e}$ comes from higher redshifts.

We then examine how tight a constraint on reionization 
redshift, $z_r$ (defined here as the redshift at which the IGM is half neutral), CMBPol can place, marginalizing
over the uncertain duration of reionization, parametrized here by $\Delta$. We contrast this constraint with one from 
existing WMAP 5
year data. The result of this calculation is shown in Figure \ref{fig:tau_reion_const}, for two different choices of input,
`true' reionization history. The two models are a low reionization redshift ($z_r = 7,\Delta=0.5$) model, and a high reionization
redshift ($z_r = 13,\Delta=0.5$) model. Both models are consistent with current WMAP data to within a $99\%$ confidence level. In 
each underlying model, the figure
illustrates that CMBPol should provide a tight constraint on the timing of reionization. The tight constraint on high
$z_r$ models may be very valuable, if the IGM is indeed reionized at high redshift. As emphasized in \S \ref{sec:obs_probes},
other observational probes are unlikely to be sensitive to high redshifts, $z \gtrsim 10-12$, soon. In this case, others probes
may first deduce that there are significant quantities of ionized gas at $z_r \gtrsim 10$ (by failing to detect neutral gas at lower redshifts), 
but will learn little else. The figure
illustrates that CMBPol can obtain much more detailed information in these high $z_r$ scenarios. If the IGM is reionized at lower redshift, other observational probes may already have definitively detected 
diffuse neutral gas in the IGM. In this case, CMBPol measurements should still 
tighten constraints and provide an independent check with a different set of systematics.



\subsubsection{\label{sec:highlowztau}Constraints on high and low redshift contributions to the optical depth}
\label{sec:tau_highz}

The improved sensitivity and resolution of CMBPol will push $E$-mode power spectrum measurements 
into the $\ell \sim 10-20$ range, and allow one to constrain more than just the total optical depth.
The next best measured quantity will be the difference between the high redshift and low redshift
contributions to the total $\tau$ \cite{MorHu08a} (e.g., the shape of the second principal component (PC)
of $x_e(z)$ in 
\S \ref{sec:pcs}).
Given that the current constraint on instantaneous reionization models 
places the redshift of reionization at $z\approx 10$ 
($z_{\rm reion}=11.0 \pm 1.4$ from WMAP5 \cite{Dunetal08}), 
it is interesting to ask how well polarization can limit the contribution to 
the total optical depth from $z<10$ and from $z>10$. 
Instantaneous models maximize the 
low-redshift contribution, so extended reionization and more exotic models 
with partial ionization at high redshift would show up as detections of 
nonzero $\tau(z>10)$ and a reduced value of $\tau(z<10)$ relative to 
the total optical depth.

The principal component analysis of \S~\ref{sec:pcs} provides one way 
to compute general bounds on the optical depth from fixed
wide bins in redshift. The optical depth due to ionization between 
redshifts $z_1$ and $z_2$ is
\begin{equation}
\tau(z_1<z<z_2) = 0.0691(1-Y_p)\Omega_b h
\int_{z_1}^{z_2} dz \frac{(1+z)^2}{H(z)/H_0}x_e(z).
\label{eq:tauz1z2}
\end{equation}
By computing $x_e(z)$ from the PC amplitudes of individual 
MCMC samples [Eq.~(\ref{eq:mmutoxe})], we can use 
eq.~(\ref{eq:tauz1z2}) to obtain constraints on the derived 
parameters $\tau(6<z<10)$ and $\tau(10<z<z_{\rm max})$.
[Since we assume full ionization at $z<6$ for all models, 
the constant contribution of $\tau(z<6)\approx 0.036$ is 
subtracted from the low redshift optical depth parameter.]

For a given choice of $z_{\rm max}$ we expect parameter constraints to 
only be accurate as long as there is no significant ionization between 
$z_{\rm max}$ and $z\sim 1100$. Nevertheless, estimates of 
$\tau(10<z<z_{\rm max})$ can be sensitive to ionization events at 
$z > z_{\rm max}$ due to the fact that high redshift ionization affects 
a wide range of multipoles that overlap the range affected by 
ionization at $z < z_{\rm max}$.  Therefore, $\tau(10<z<z_{\rm max})$ 
can effectively be thought of as $\tau(z > 10)$ \cite{MorHu08a}.  
If future data show signs of high redshift ionization, then it will be 
necessary to check the robustness of constraints by increasing the 
chosen value of $z_{\rm max}$ to ensure accurate parameter estimates.

The constraints on the optical depth coming from low redshift, $\tau(6<z<10)$, 
and from high redshift, $\tau(z>10)$, are plotted in 
Fig.~\ref{fig:highlowztau} for current data and for the 
expected sensitivity of Planck and cosmic variance limited CMBPol. 
The error forecasts assume the same fiducial ionization history 
and cosmology as for Fig.~\ref{fig:pcs}, which has roughly equal 
optical depths of $\sim 0.02$ coming from $6<z<10$ and $z>10$.  

The contours in the left hand panel of Fig.~\ref{fig:highlowztau} are elongated in the 
direction of constant $\tau$, showing that the total optical depth 
is much better constrained than its contributions from 
more restricted redshift ranges even in the cosmic variance limit.
The contours are cut off at the upper edge of the plot 
in Fig.~\ref{fig:highlowztau} due to the upper limit 
$\tau(6<z<10) \leq 0.039$, which comes from requiring that the ionized 
fraction be no more than unity over this range of redshifts.
While the maximum likelihood models from the Planck and CMBPol 
MCMC analyses coincide with the fiducial values plotted as the cross 
in Fig.~\ref{fig:highlowztau}, the confidence regions are shifted 
slightly in the direction of higher optical depth coming from $z<10$.
This may be partly due to the fact that $6<z<10$ is a fairly narrow 
redshift range compared to the full extent of the principal components, 
which we have taken to be $6<z<30$ here. Consequently, some of the
high variance PCs that can safely be neglected when considering 
$C_{\ell}^{\rm EE}$ or total $\tau$ due to having many oscillations 
in redshift are much smoother over the range $6<z<10$. By fixing these 
components to have zero amplitude, we are ignoring any potential 
perturbations they can induce in $\tau(6<z<10)$.
%
%
The maximum perturbation to $\tau(6<z<10)$ from the next principal component 
($\mu = 6$) is $\sim 0.005$, comparable to the 68\% uncertainty in this 
parameter with cosmic variance limited data. However, reionization histories 
with such large values of the higher-variance PCs are extreme and possibly 
even unphysical, so in practice the effects of the components that we ignore
on the constraints in Fig.~\ref{fig:highlowztau} should be reasonably small.


\begin{figure}[t]
\psfig{file=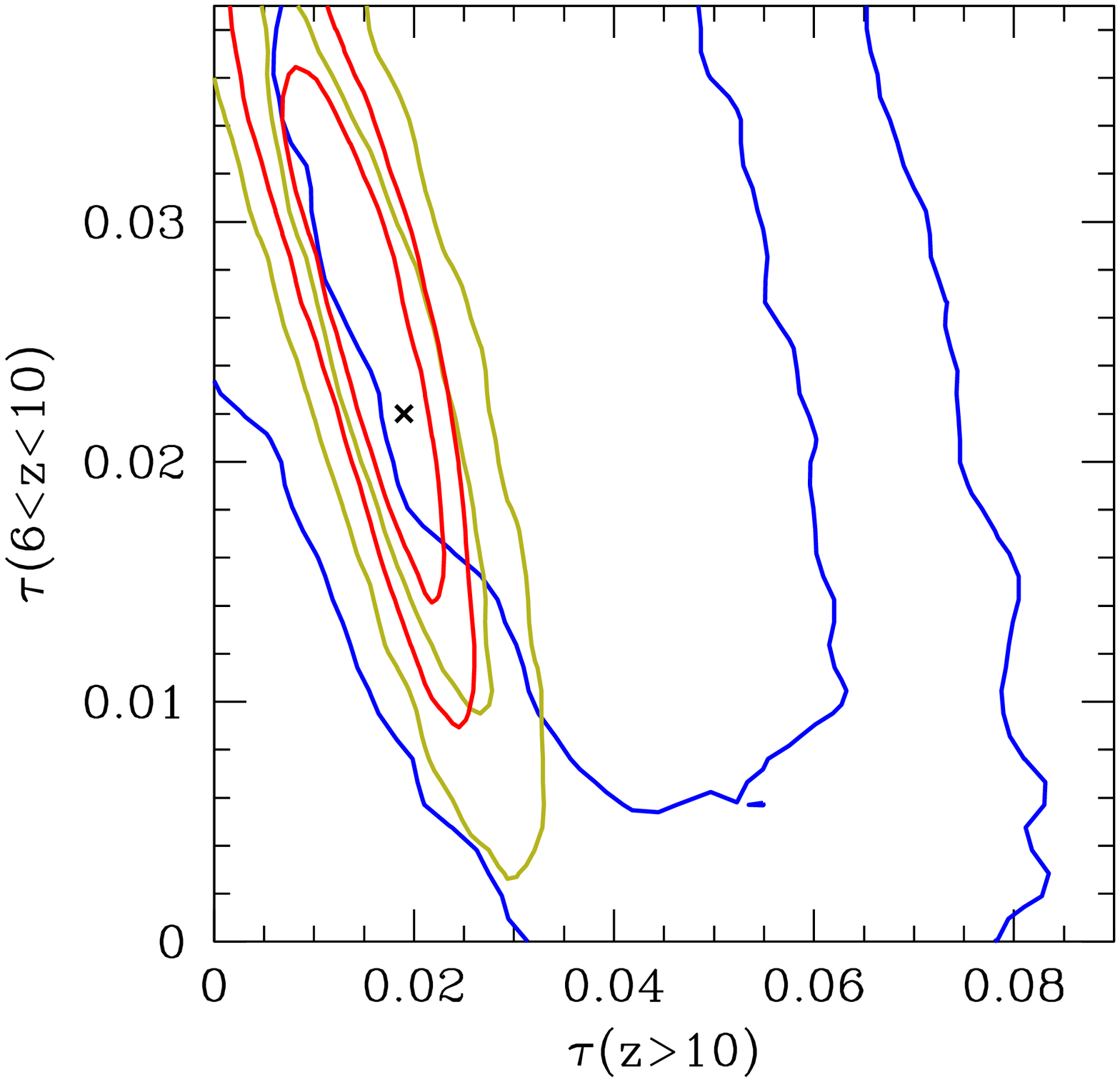, width=3.5in}
\psfig{file=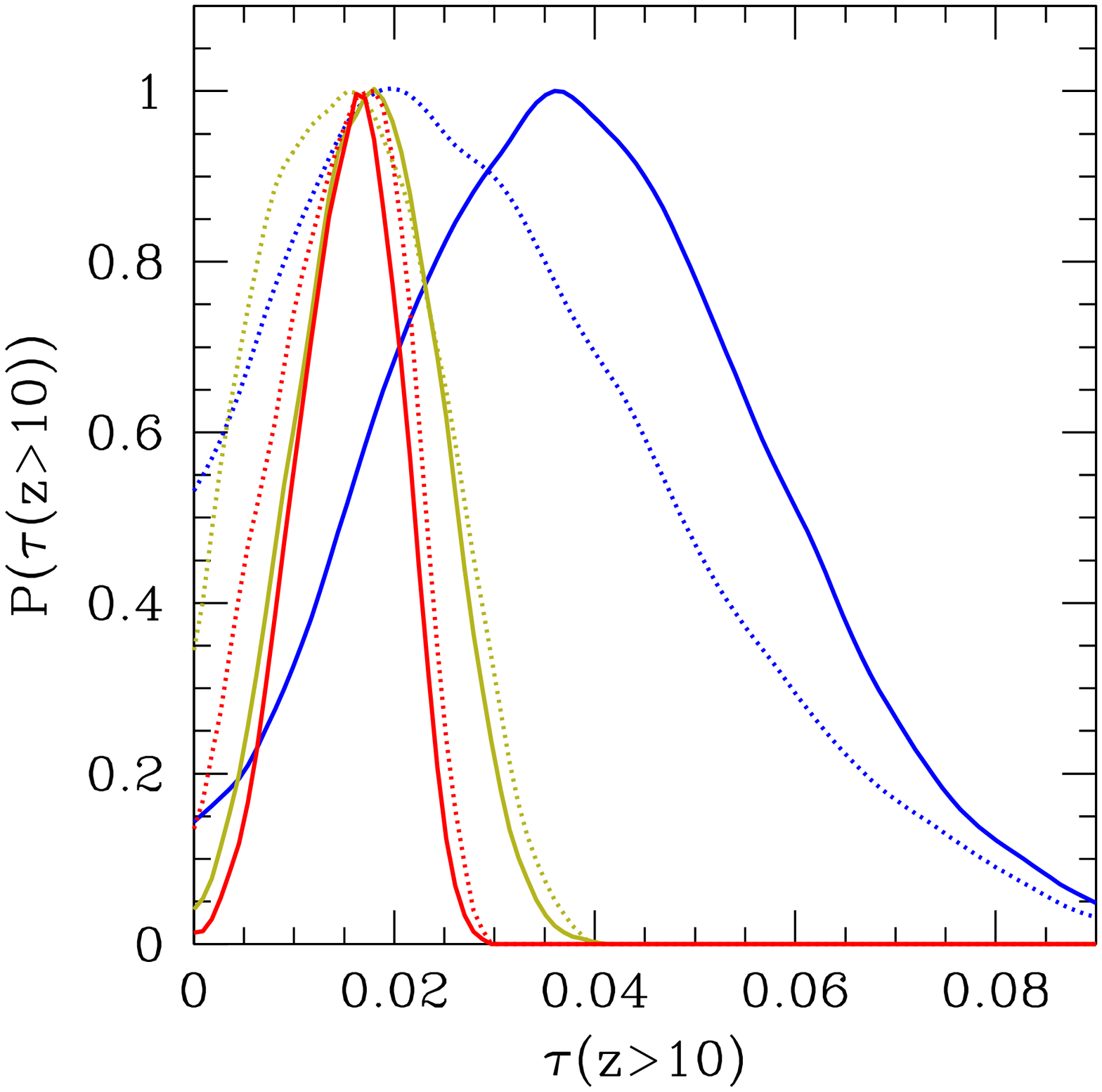, width=3.5in}
\caption{Constraints on contributions to the reionization optical 
depth from $6<z<10$ and $z>10$ with temperature and polarization data 
from WMAP5 (\emph{blue}) and forecasted for large scale polarization 
from Planck (\emph{gold}) and CMBPol (\emph{red}). 
\emph{Left}: 68\% and 95\% CL 2D contours computed 
using the same set of MCMC samples as in Fig.~\ref{fig:pcs}. 
The values of $\tau(6<z<10)$ and $\tau(z>10)$ for the fiducial 
ionization history are indicated by a cross.
\emph{Right}: Marginalized posterior probability (\emph{solid}) and 
mean likelihood (\emph{dotted}) of $\tau(z>10)$. The discrepancy between 
the posterior distribution and mean likelihood for WMAP and for 
Planck at low $\tau(z>10)$ is due to the priors set by the range of 
possible values of the optical depth from $6<z<10$ and $z>10$. Only for CMBPol 
is there both a small number of MCMC samples and a small likelihood at 
$\tau(z>10)=0$.}
\vskip 0.25cm
\label{fig:highlowztau}
\end{figure}

The key point illustrated in Figure \ref{fig:highlowztau} is that only CMBPol can strongly reject a significant
high redshift contribution to the total optical depth. A strong limit on the high redshift contribution to the
optical depth would rule out the presence of an early and efficient mode of star formation, and constrain the possible
contribution of ionizing photons produced by annihilating or decaying dark matter. Alternatively, detecting a
significant high redshift contribution would point to an exotic source of ionizing photons.

\subsubsection{Constraints on the duration of reionization}
\label{section:zr_dz}

Next we consider whether large scale polarization measurements can constrain
the {\em overall duration} of the EoR, in addition to the redshift at which the
IGM becomes half-neutral. 

We use the simple two parameter model described in section \ref{sec:tau_implications}.
We explore two sets of parameters: 
\begin{itemize}
\item[(i)] ``Low $\tau$ model,''
($z_r$,$\Delta$)=(10.5, 2.0), whose optical depth, $\tau=0.088$, is close
to the WMAP best-fitting value, $\tau_{WMAP}=0.087\pm 0.017$ \cite{Dunetal08},
\item[(ii)]  ``High $\tau$ model,''
($z_r$,$\Delta$)=(13.0, 3.5), whose optical depth, $\tau=0.119$, is close
to the upper 95\% CL limit of the WMAP  value.
\end{itemize}

Figures~\ref{fig:tanh1} and \ref{fig:tanh3} show the
WMAP 5-year constraints (blue) and forecasts (yellow for Planck and
orange for CMBPol) for the low and high optical depth scenarios. 
In the low $\tau$ model although Planck and CMBPol tightly constrains $z_r$ (as in \S \ref{sec:tau_implications}),
the duration of reionization, $\Delta$, is essentially unconstrained. Even the unphysical case of an instantaneous
reionization model lies within the $1-\sigma$ error contour. Models with large $\Delta$ are either within the $2-\sigma$ error
contours, or are already excluded by 
$z \sim 6$ quasar spectra (see \S \ref{sec:quasars}). This is partly because the ionization history in this model
is symmetric around $z_r$. For instance, our current parametrization does not allow models where the ionization history rises abruptly for
$z \leq z_r$, but is more gradual above $z \geq z_r$. In this case, one can constrain the high redshift 
contribution to the opacity,
as we demonstrated in \S \ref{sec:tau_highz}. In our high $\tau$ input model, one expects a weak constraint on the 
duration parameter,
$\Delta$, from CMBPol. In this case, the instantaneous reionization and very rapid reionization models can be 
rejected at $2-\sigma$ by
CMBPol, although Planck can not distinguish such models. Unfortunately,
it is hard to distinguish between more probable models with, e.g., $\Delta=2$ and $\Delta=4$.

\begin{figure}
\centering \noindent
\includegraphics[width=12cm]{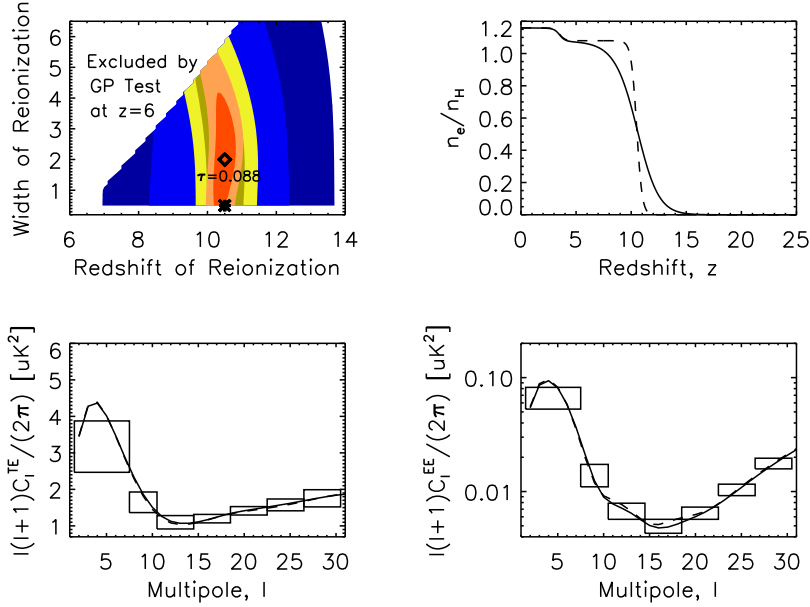}
\caption{%
Low $\tau$ model, $z_r=10.5$ and $\Delta=2.0$. 
Planck or CMBPol cannot distinguish between the extended and
 instantaneous reionization scenarios  for this case.
({\it Top Left}) 
Projected constraints on the epoch and duration of reionization,
 parametrized by the tanh model given in Eq.~(\ref{eq:tanh}).
The blue contours show the existing WMAP 5-year constraints at the 68\%
 and 95\% CL, while the yellow and orange contours show the projected
 constraints from Planck and CMBPol, respectively. The diamond and star
 show the input model and an instantaneous reionization model with $\Delta=0.5$,
 respectively. ({\it Top Right}) 
Reionization history. The solid line shows the input model (diamond in
 the top-left panel), while the dashed line shows an instantaneous
 reionization model (star in the top-left panel). 
({\it Bottom Left}) Temperature-$E$ polarization cross power spectrum. The
 boxes show the expected 68\% CL bounds from CMBPol. The solid and
 dashed lines correspond to those in the top-right panel. 
({\it Bottom Right}) $E$ polarization power spectrum. The
 boxes show the expected 68\% CL bounds from CMBPol. The solid and
 dashed lines correspond to those in the top-right panel. 
}
\label{fig:tanh1}
\end{figure} 

\begin{figure}
 \centering \noindent
\includegraphics[width=12cm]{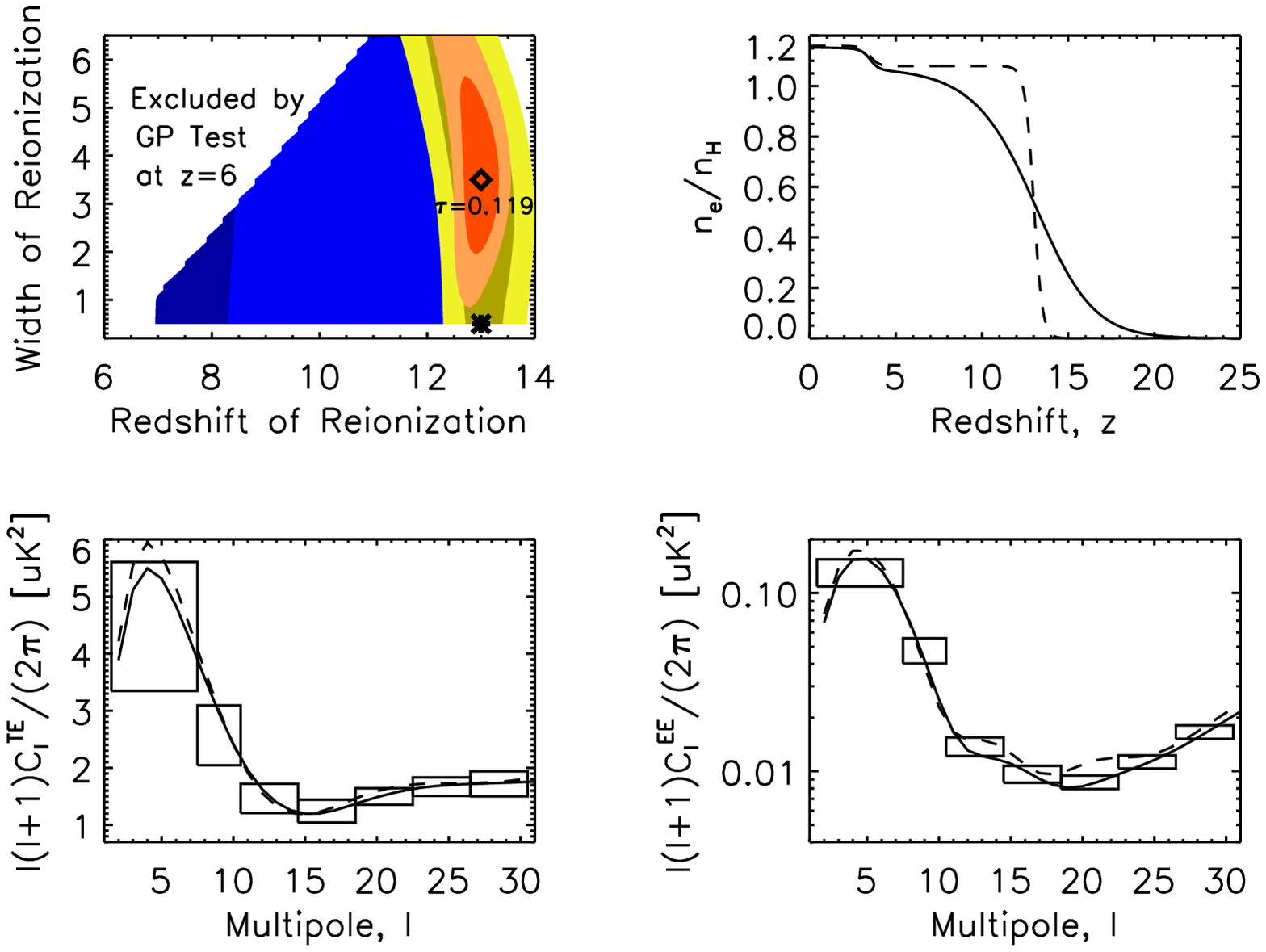}
\caption{%
Same as Fig.~\ref{fig:tanh1}, but for the high $\tau$ model, $z_r=13.0$
 and $\Delta=3.5$. Although Planck cannot distinguish between the extended and
 instantaneous reionization scenarios, CMBPol can reject the
 instantaneous reionization at more than 95\% CL. These models are
 discriminated best at $l\sim 20$.}
\label{fig:tanh3}
\end{figure}



\subsubsection{Constraints on a physical model}
\label{section:physical_model}

The `$z_r-\Delta$' parametrization of the previous subsection is convenient, but somewhat arbitrary,
and so we would like to explore an additional model to test whether this impacts our main conclusions. 
In this subsection we consider an alternate two parameter model for the reionization history.
This model is also somewhat simplistic, but in combination with 
our previous model and the principle component analysis of \S \ref{sec:pcs}, it should provide a good sense for
CMBPol's capabilities.

In the model of this section, the ionization fraction is directly related 
to the fraction of mass that is in dark matter halos above some minimum mass. This form is motivated
by assuming that the star formation rate is proportional to the rate at which matter collapses
into galactic host halos.
See e.g. \cite{Alvarez06,Furlanetto06a} for a discussion of this model. 
More specifically, we assume that
\begin{equation}
x_e(z) = 1 - {\rm \exp}\left(-\zeta \ {\rm erfc}\left[\frac{\delta_c(z)}{\sqrt{2 \sigma^2(M_{\rm min})}}\right]\right).
\label{eq:x_phys}
\end{equation}
Here the error function term, is the 
collapse fraction for halos with mass above $M_{\rm min}$ according to Press-Schechter theory,
$\delta_c(z) = 1.686/D_L(z)$ is the linear theory overdensity for spherical top-hat collapse at redshift $z$, 
extrapolated to today, $D_L(z)$ is the linear growth factor, 
and $\sigma^2(M_{\rm min})$ is the variance of the linear density field
extrapolated to today, and smoothed on a mass scale $M_{\rm min}$. We treat $\zeta$ and $M_{\rm min}$ as free parameters.
The parameter $\zeta$ is an efficiency parameter that depends physically on the number of ionizing photons produced
per stellar baryon, the fraction of gas converted into stars, the fraction of ionizing photons that 
escape from host halos and make it into the IGM, as well as the average number of recombinations
during reionization. This expression is similar to the form $x_e(z) = \zeta f_{\rm coll}(z)$ 
(with $f_{\rm coll}$ denoting the collapse fraction in halos above $M_{\rm min}$) that is
sometimes used, but our form goes smoothly to $x_e = 1$ as $\zeta f_{\rm coll}$ becomes large.
The parameter $M_{\rm min}$ is the minimum mass for a halo to
host an ionizing source. In reality $\zeta$ and $M_{\rm min}$ likely vary with redshift and environment, but we neglect
this here in order to adopt a simple two parameter model.

We computed a physical model which is tuned to have the same optical depth $\tau=0.12$ as the high $\tau$ model in the previous section; it turns out to have a roughly comparable spread in the reionization epoch as well. Fig. \ref{fig:tanh_phys_comparison} shows that CMBPol can distinguish between the physical
model and an instantaneous reionization  model with the same $\tau$ at $\sim 95$\% CL. As in the $z_r-\Delta$ parametrization, it is hard 
to distinguish models that have the same $\tau$, yet different durations: only the instantaneous model can be rejected. This is seen in the contour
plot as a degeneracy
between very efficient yet rare source models (high $\zeta$, high $M_{\rm min}$) where reionization occurs rapidly, and low efficiency 
abundant source models (low $\zeta$, low $M_{\rm min}$) in which reionization is
gradual. 
This supports the basic conclusion of the previous section: 
CMBPol can reject very rapid reionization models, but detailed constraints on the duration of reionization are impossible for smoothly varying
reionization models.  
 
\begin{figure}
\centering \noindent
\includegraphics[width=12cm]{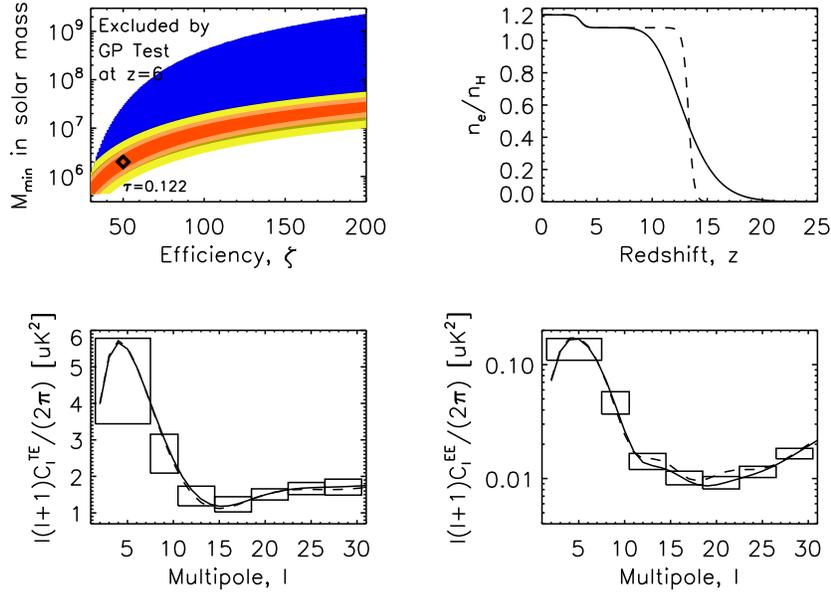}
\caption{Same as Fig. \ref{fig:tanh3}, but for
physical model with the minimum mass of $M_{min}=2\times 10^6~M_{\odot}$ and
efficiency of $\zeta=50$. While CMBPol cannot distinguish between different physical models
along the direction of having the same $\tau$ value, it can distinguish between the physical
model and an instantaneous reionization  model with the same $\tau$ at $\sim 95$\% CL.}
\label{fig:tanh_phys_comparison}
\end{figure}

\subsubsection{{\label{sec:pcs}}Constraints on Principal components}

\begin{figure}[h]
\centerline{\psfig{file=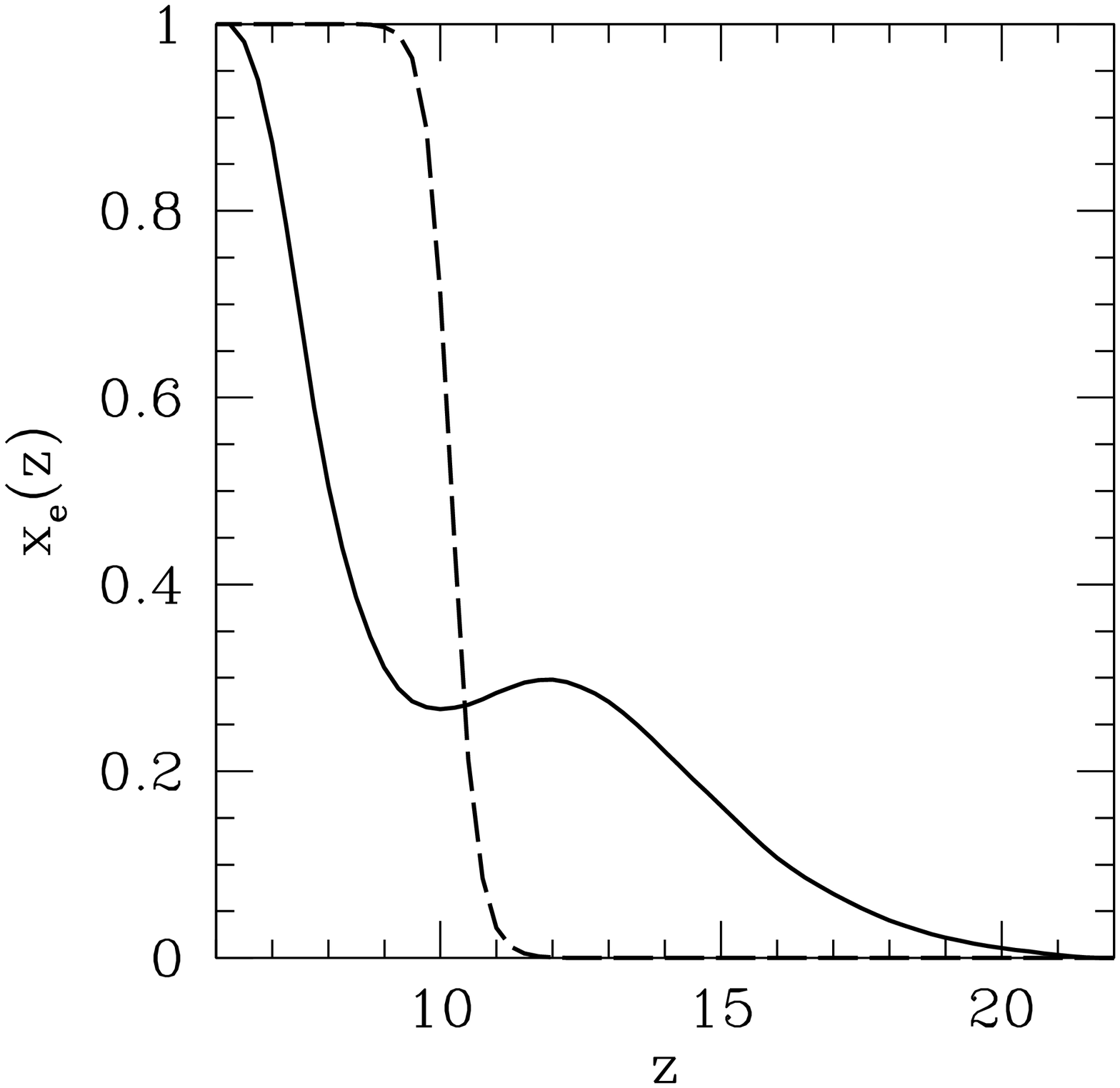, width=3.5in}
\psfig{file=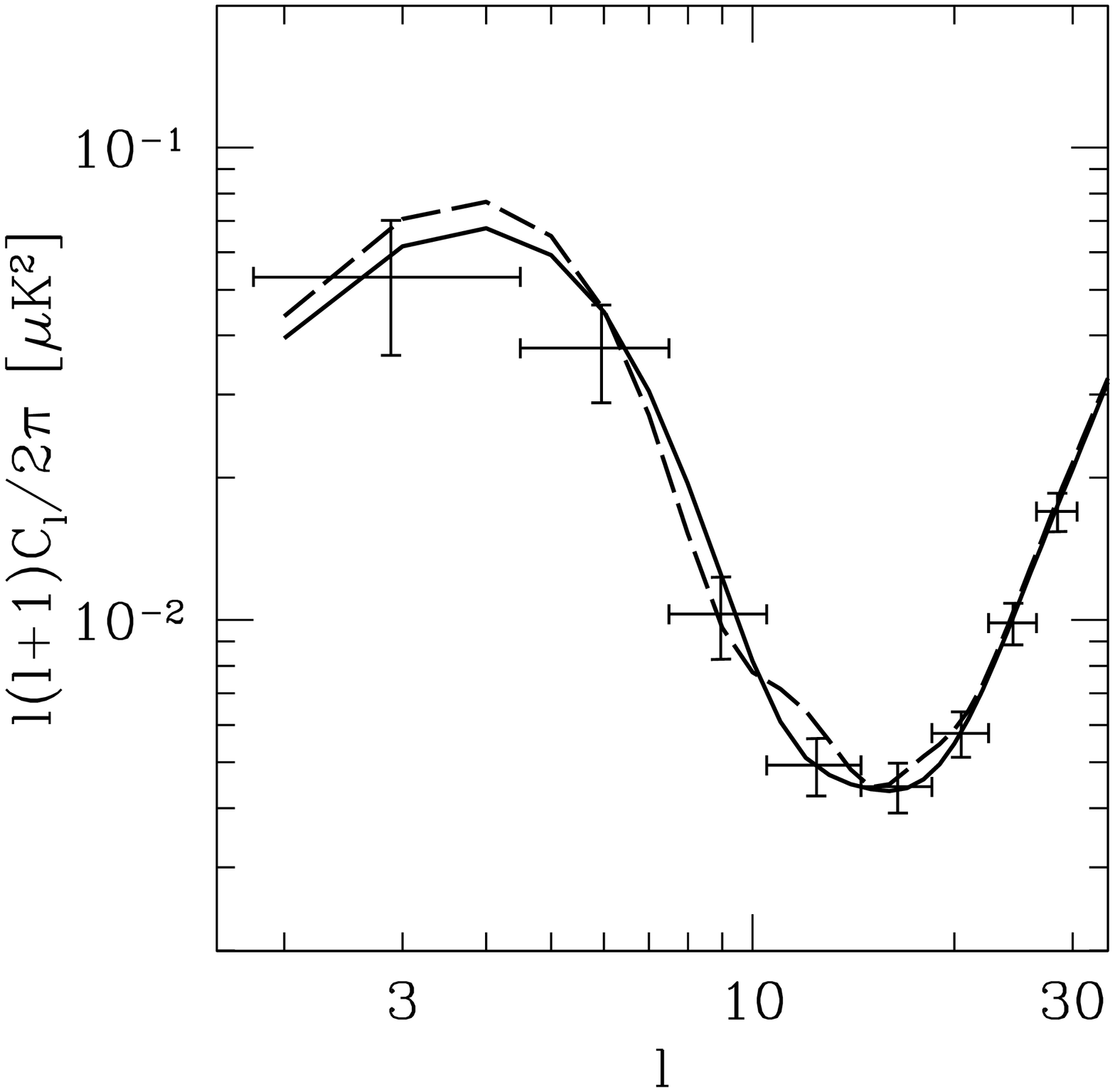, width=3.5in}}
\caption{\emph{Left}: Ionization histories for the fiducial model used 
for MCMC with principal components (\emph{solid}) and for an 
instantaneous reionization model with the same total optical depth, 
$\tau=0.077$ (\emph{dashed}).
\emph{Right}: $E$-mode polarization power spectra for the ionization 
histories in the left panel. The 68\% CL errors for a cosmic variance limited 
CMBPol experiment are plotted for the fiducial model (binned in $\ell$).
}
\vskip 0.25cm
\label{fig:xeclee}
\end{figure}

The true reionization history may be more complex than the parametrized models of the previous two sub-sections.
For example, feedback effects from the first generation of sources will likely impact the formation of subsequent
galaxies and/or quasars, forcing the efficiency parameter and minimum host mass to vary with redshift and environment.
If this is the case, one can hope to learn more about the reionization history than implied by the previous two
sub-sections. In addition to the timing of reionization, and constraints on the high redshift contribution to the opacity,
one may hope to learn further information regarding the duration of the reionization process. For instance, is the reionization
process even monotonic?

As an example of a more general reionization history, we consider a double-peaked reionization 
history from Furlanetto \& Loeb \cite{FurLoe05}. These authors investigate the conditions
under which double-peaked reionization histories are physically plausible, and find that such histories require fine-tuning
and are somewhat difficult to arrange, but not impossible. We hence adopt their model with a clumping factor of $C=3$ and a high
virial temperature due to photoionization heating of $T_h = 2.5 \times 10^5$ K. The ionization history in this 
model, its $E$-mode power spectrum with CMBPol error
estimates, and
the same for an instantaneous reionization model with the same $\tau$ ($\tau=0.077$) are shown in Figure \ref{fig:xeclee}.
 The instantaneous and double-peaked model can be distinguished
at high significance (in excess of 3-$\sigma$). This illustrates that, if more freedom is allowed in the underlying reionization history
than in the previous sub-sections, one can constrain more than just the total optical depth.

This motivates a more conservative approach that complements the constraints from the previous sub-sections.
Here we allow $x_e(z)$ to be a free function of redshift and see what constraints the data place on the
form of this function with minimal theoretical assumptions.
One simple implementation of this approach is 
to parametrize the ionization history using the values of $x_e$ in 
several wide redshift bins \cite{LewWelBat06,ColPie08}.
An alternative parametrization that we employ here is 
principal components (PCs) of the ionization history, an orthogonal 
set of basis functions for $x_e(z)$ ranked in order of how well they 
can be measured with large scale polarization data \cite{HuHol03}. 
The interpretation of constraints on PCs is less intuitive than for 
the binned ionization fraction, but the advantages of PCs are that they are 
largely decorrelated from one another, there is less need for arbitrary 
redshift bin choices, and the impact of reionization on the CMB 
polarization is concentrated within a small number of the 
best measured components.

The principal components $S_{\mu}(z)$ are defined as the 
eigenfunctions of the Fisher matrix that describes the information about 
reionization contained in the CMB polarization power spectrum on 
large scales \cite{HuHol03},
\begin{eqnarray}
F_{ij} & = & \sum_{\ell=2}^{\ell_{\rm max}}\left(\ell+\frac{1}{2}\right)
 \frac{\partial \ln C_{\ell}^{\rm EE}}{\partial x_e(z_i)}
 \frac{\partial \ln C_{\ell}^{\rm EE}}{\partial x_e(z_j)}, \\
& = & (N_z+1)^{-2}\sum_{\mu=1}^{N_z}
 S_{\mu}(z_i)\sigma^{-2}_{\mu}S_{\mu}(z_j), \nonumber
\label{eq:pcfisher}
\end{eqnarray}
where $\ell_{\rm max} \sim 50-100$ is large enough to include all effects of 
the mean ionization fraction, and the factor $(N_z+1)^{-2}$ is included 
so that the eigenfunction shapes are independent of the initial 
redshift bin width $\Delta z = z_{i+1}-z_i$ as $\Delta z \to 0$.
The ordering of the PCs is based on their expected variances, given by 
the inverse eigenvalues of $F_{ij}$, $\sigma^2_{\mu}$, 
with the best constrained PCs having the lowest values of 
$\sigma_{\mu}$.
For the set of PCs used here, we compute $F_{ij}$ at a fiducial model 
of constant $x_e^{\rm fid}=0.15$ in redshift bins with spacing $\Delta z=0.25$ 
over a redshift range $6<z<30$.  
The first five of these components are plotted on the left side 
of Fig.~\ref{fig:pcs}. 
The minimum redshift $z_{\rm min}=6$ is motivated by observations of quasar 
absorption spectra. 
For details on how results depend on the chosen fiducial model and 
maximum redshift, $z_{\rm max}$, see Ref.~\cite{MorHu08a}.
The above equation assumes full sky coverage and
sensitivity limited only by cosmic variance, but these choices have little 
effect on the PC shapes.

\begin{figure}[t]
\centerline{\psfig{file=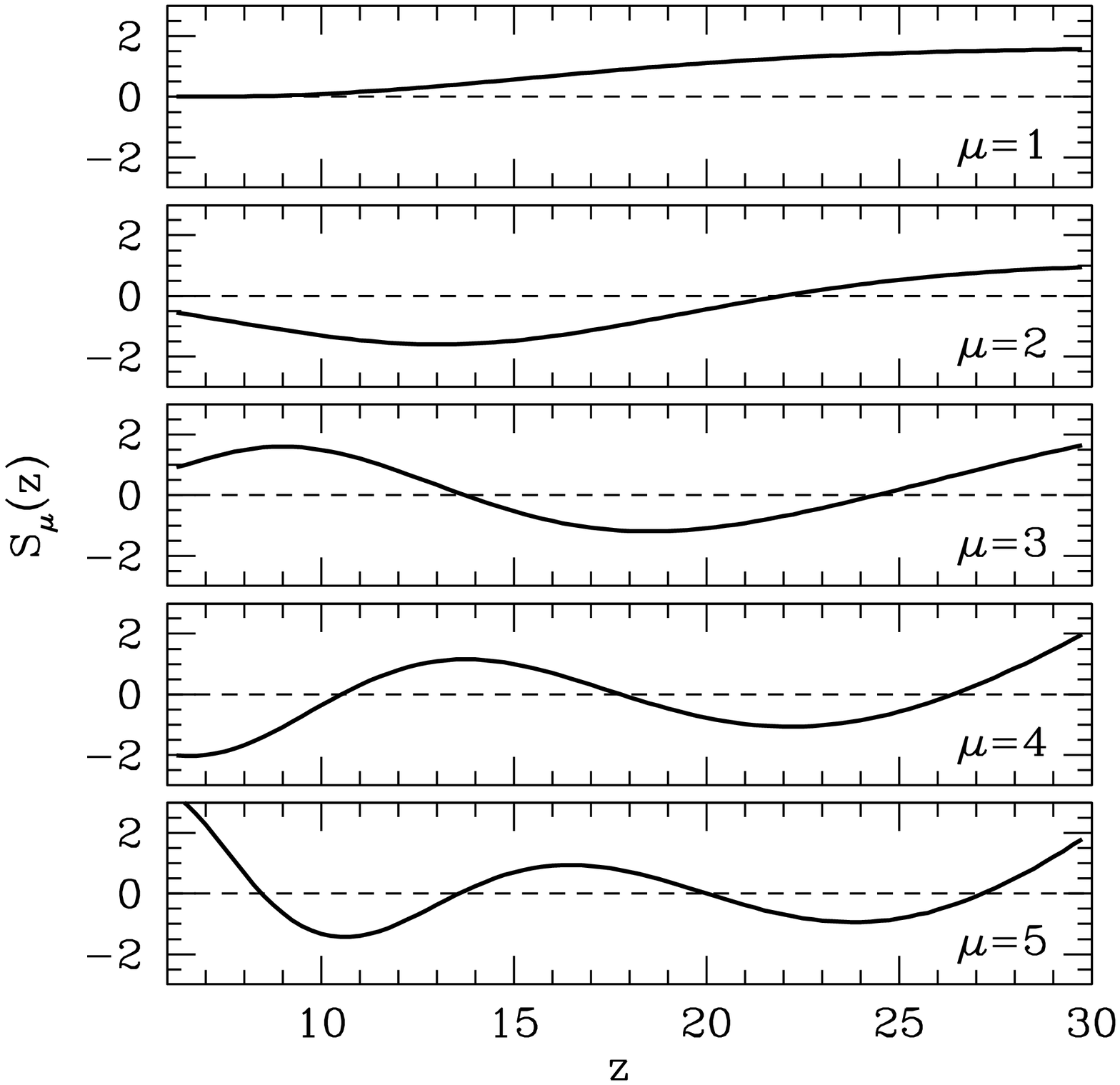, width=3.in}
\psfig{file=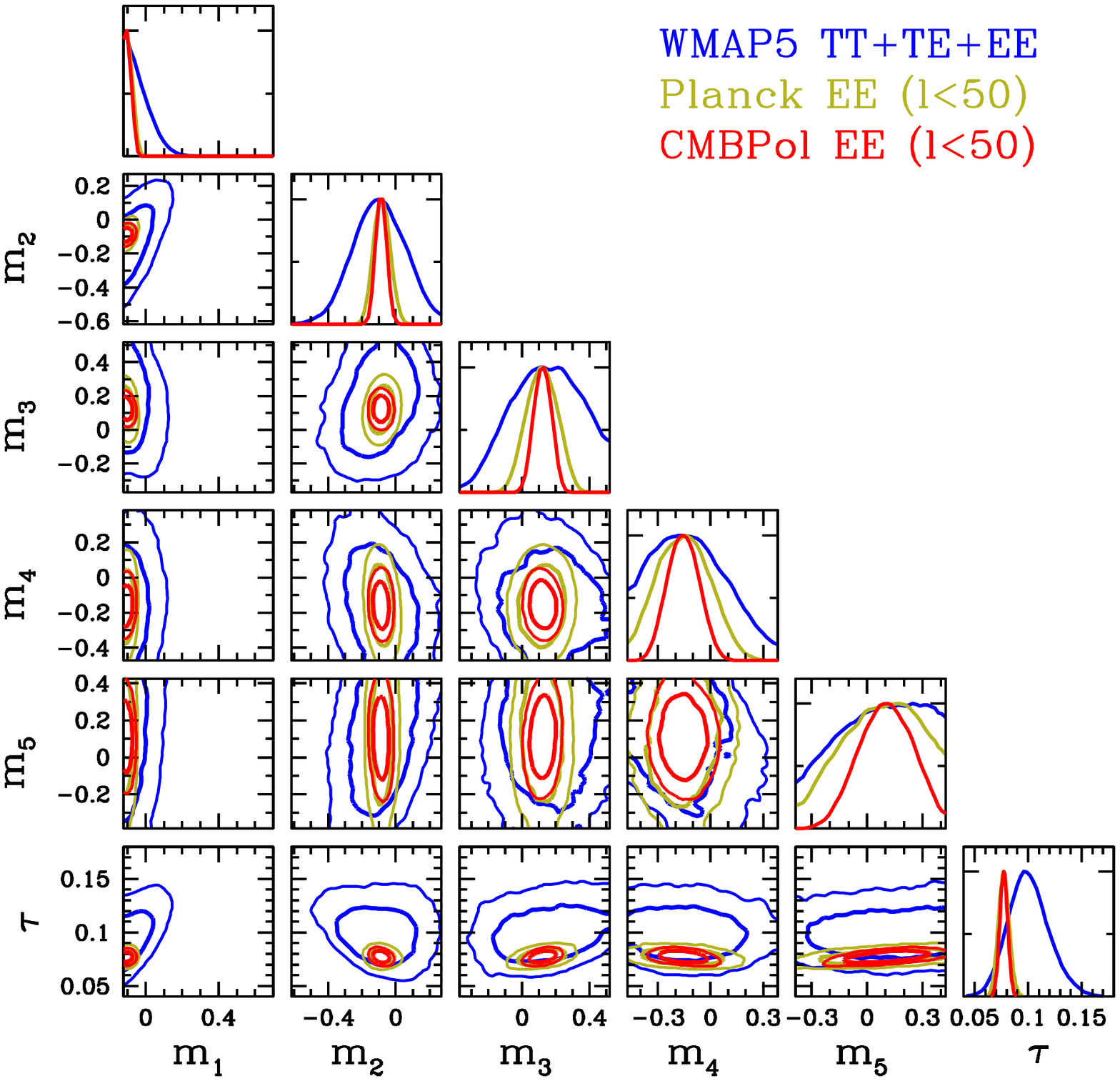, width=4.in}}
\caption{\emph{Left}: Five lowest variance principal components 
of the ionization history for large scale $E$-mode polarization. 
\emph{Right}: Constraints on principal components 
and total optical depth to reionization from 5-year WMAP data 
including the full temperature and polarization power spectra 
(\emph{blue}), and forecasts for Planck (\emph{gold}) and a 
cosmic variance limited CMBPol experiment (\emph{red}) using 
polarization data at $\ell\leq 50$. Each set of contours shows 
68\% and 95\% CL regions.
}
\vskip 0.25cm
\label{fig:pcs}
\end{figure}

Any function of redshift between $z_{\rm min}$ and $z_{\rm max}$ can 
be expressed as a linear combination of PCs,
\begin{equation}
x_e(z)=x_e^{\rm fid}(z)+\sum_{\mu=1}^N m_{\mu}S_{\mu}(z).
\label{eq:mmutoxe}
\end{equation}
Only the $3-5$ lowest variance PCs affect the polarization power spectrum 
to a significant degree (i.e. larger than or comparable to 
cosmic variance) \cite{HuHol03,MorHu08a}. 
One can therefore truncate the sum in Eq.~(\ref{eq:mmutoxe}) when 
estimating parameters from CMB data with $N=3-5$ while still retaining the 
full information about reionization present in the large scale polarization.
For the results presented here and in \S~\ref{sec:highlowztau}, 
we always use the five components plotted
in Fig.~\ref{fig:pcs} as parameters for MCMC analysis.

The right panel of Fig.~\ref{fig:pcs} shows MCMC constraints on 
five PCs and the total optical depth $\tau$ from current and future CMB data.
The largest set of contours are from five-year WMAP temperature and 
polarization power spectra, using MCMC estimates of the five 
reionization parameters in addition to the baryon density, CDM density, 
Hubble constant, and scalar amplitude and tilt \cite{MorHu08c}. 
Forecasted constraints on the five PCs from polarization at $\ell\leq 50$ 
are plotted for Planck, assuming a sensitivity of 
$w_p^{-1/2}=63.1~\mu$K~arcmin and beam size $\theta_{\rm FWHM}=7$~arcmin, and 
for a high resolution, cosmic variance limited experiment (``CMBPol''). 
The fraction of sky usable for cosmological constraints is assumed to be 
$f_{\rm sky}=0.8$ for both future experiments.
We include the temperature data for WMAP since it provides some 
additional constraints to the current polarization 
data, but for Planck and CMBPol essentially all of the information 
about the mean ionization history can be obtained from the large scale 
polarization \cite{MorHu08a,MorHu08c}. To find the projected Planck and CMBPol 
parameter uncertainties we include polarization data up to $\ell=50$ since 
treating the ionization history as a free function up 
to high redshift allows for significant variation in $C_{\ell}^{\rm EE}$ 
at smaller scales than in typical parametrizations of reionization.
For these forecasts, the amplitude of the temperature and polarization 
spectra at high $\ell$ is held fixed to the best fit to current data by 
keeping the value of $A_s e^{-2\tau}$ fixed for each MCMC sample.

While the WMAP constraints in Fig.~\ref{fig:pcs} come from real data and 
therefore reflect the true reionization history, we must choose the 
model for the simulated data used in the Planck and CMBPol forecasts. 
We take the fiducial ionization history to be the double-reionization model of Figure \ref{fig:xeclee}
and we compute MCMC likelihoods relative 
to the $E$-mode polarization spectrum of this model. 
This fiducial model has a total optical depth of $\tau=0.077$ and 
a period of reionization extending from $z\approx 20$ to $z\approx 7$.
The estimates of $\tau$ and the principal component amplitudes 
recovered from the MCMC analysis agree well with the true parameter values 
of this input model.

The boundaries of plotted regions for the principal components on 
the right side of Fig.~\ref{fig:pcs} 
are the edges of our priors on individual PCs.  All potentially 
physical ionization histories (i.e. $0\leq x_e \lesssim 1$) are 
included within these priors, including those that appear to violate 
the allowed range of $x_e$ when the sum of PCs is truncated \cite{MorHu08a}. 
However, this does not mean that every model 
inside these bounds is physically allowed or plausible. 
The PC priors are therefore in the spirit of the conservative approach 
to reionization constraints in which we allow for the largest 
conceivable variation in the ionization history.

If we allow the priors on PC amplitudes to define the range over which 
measurements enable interesting comparisons between models, then 
we find that current data places tight constraints on the first 
principal component, roughly corresponding to $\tau$, and weakly limits 
the allowed range of the second component. Planck and CMBPol are 
expected to provide accurate estimates of several components: 
about $3-4$ for Planck and $4-5$ for cosmic variance limited CMBPol.
Improvements in $\tau$ and the two lowest variance components between 
Planck and CMBPol are minimal, with most of the impact of having 
truly cosmic variance limited data appearing in the higher PCs.

It is important to note, however, that principal component parameters 
estimated at a significant level compared with the conservative 
$0\leq x_e\leq 1$ priors do not necessarily lead to interesting constraints 
on physically motivated models. 
For many models of reionization that are currently thought to be realistic, 
the variation between models in higher variance PCs that undergo several 
oscillations in redshift is likely to be negligible. 
Therefore, the number of PCs that can be measured 
to an ``interesting'' level of precision depends on the range of possible 
variation in each of the components for the set of models being considered.
Although a cosmic variance limited experiment is expected to measure 
$4-5$ components at a significant level compared with the full 
possible range when the ionization fraction is allowed to be a free 
function of redshift, under more restricted classes of models like those 
in \S~\ref{section:physical_model}, it may
be difficult to measure even two parameters at high significance.

Until observational limits on reionization and our theoretical understanding 
greatly improve our confidence in the specific behavior of the 
ionization fraction over time, it is worth keeping in mind that 
the true reionization scenario might be stranger than what we expect.
Forecasts for models with a solid physical basis provide a realistic 
view of the advances we can expect from one experiment to the next, 
while the projected uncertainties on parameters like principal components 
that form a more model-independent description of the ionization history
place an upper limit on the information about reionization that 
can be extracted from the CMB power spectra on large scales.

\section{\label{sec:otherparams}Impact of improved knowledge of reionization on other parameters}


The $\tau$ determination for Planck will not be significantly
degenerate with any other parameter apart from the amplitude of scalar
perturbations and in some instances with the scalar--to--tensor ratio
$r$ \cite{ColPie08}.  This is also true when a reconstruction of the
reionization history from the data is attempted, however
discrimination between specific reionization scenarios may be a
challenge for Planck \cite{MLidd08}.

As for degeneracies, in such experiments $\tau$ would only be
significantly degenerate with the amplitude of scalar perturbation
$A_s$ and as a consequence with $\sigma_8$(see Fig.\ref{fig:twod}).
Even in non--minimal scenarios where an arbitrary scalar running spectral
index and an free tensor spectral index are considered, $\tau$ does
not show significant degeneracy with these parameters (Fig.\ref{fig:twod_run}).

\begin{figure*}
\centerline{
\includegraphics[width=10.cm]{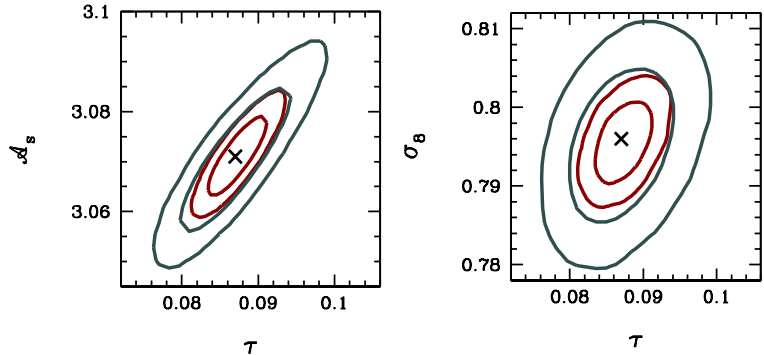}
}
\caption{Joint confidence regions, at $68 \%$ and $95 \%$ CL, in the
$\tau$--${\cal A_s}$ and $\tau$--$\sigma_8$, assuming an instantaneous
reionization. Curves are plotted for Planck (green lines) and CMBPol
(red). There are no significant degeneracies between $\tau$ and the
remaining parameters.}
\label{fig:twod}
\end{figure*}

\begin{figure*}
\centerline{
\includegraphics[width=10.cm]{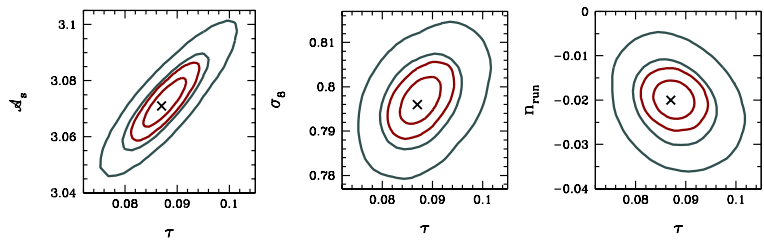}
}
\caption{Same as Figure~\ref{fig:twod} but for a fiducial model with a 
non--zero running of the scalar spectral index, $n_{\rm run}$ and a 
non--zero tensor spectral index, $n_t$.}
\label{fig:twod_run}
\end{figure*}

If CMBPol is able to measure the $B$-mode polarization signature of 
gravitational waves from inflation, accurate estimates of reionization 
parameters will be necessary to avoid biasing constraints on 
inflationary parameters. Unless the amplitude of tensor perturbations is 
close to the current upper limit set by the CMB temperature 
data, the ability to probe inflation is likely to 
rely heavily on the reionization peak in the large scale 
$B$-mode tensor spectrum since instrumental noise, angular resolution, 
and the conversion of $E$-modes into $B$-modes via lensing all present 
significant challenges for measuring $C_{\ell}^{\rm BB}$ on smaller scales. 
Without sufficient information about the reionization 
history from large scale $E$-mode polarization, estimates of inflationary 
parameters such as the tensor-to-scalar ratio $r$ and tensor tilt $n_t$ 
that make use of the reionization peak in $C_{\ell}^{\rm BB}$ 
are strongly correlated with the total optical depth to reionization. 
Accurate constraints on $\tau$ and on the shape of the reionization history 
from cosmic variance limited measurements of $C_{\ell}^{\rm EE}$ 
would break this degeneracy, enabling the use of the large scale tensor 
contribution to $C_{\ell}^{\rm BB}$ for testing models of 
inflation \cite{MorHu08b}.



\section{Probing reionization with secondary anisotropies} 
\label{sec:ss_cmb}

\subsection{Small angular scale measurements}


On scales comparable to a tenth of a degree or smaller ($\ell \geq 2000$), a second reionization observable comes into play, anisotropy created by the inhomogeneous distribution of ionized hydrogen (HII) surrounding the first sources. In this section we will discuss this probe and its potential to constrain not only the mean reionization redshift, but also its 
duration. Whether CMBPol or future ground based experiments will be the best platforms to do this measurement is currently unclear. A polarization mission which attempts to de-lens the polarization signal to reduce the lensing B-mode contamination requires high angular resolution and high sensitivity. Such a mission, if also sensitive to temperature anisotropies, would be ideal to study the secondary anisotropies generated during reionization. 

There are two observables sourced by the inhomogeneous distribution of ionized hydrogen during the EoR. First note that anisotropies due to varying integrated optical depth $\tau(\n)$ lead to anisotropies in the observed power in different directions on the sky. Because in typical reionization models the sightline fluctuations are only of order $1\%$, the contribution to the power spectrum is of order $10^{-4}$, which is below cosmic variance. \footnote{Even if one attempted to measure this contribution in one broad band power.}

The dominant small scale CMB reionization observable comes instead from the fact that the extended ionized bubbles (which are about $30 \, \mpch$ across at the intermediate EoR stages, leading to typical signatures of size $10'$ on the sky) are embedded in large scale line-of-sight velocity flows. While the standard Ostriker-Vishniac/kinetic Sunyaev-Zel'dovich (SZ) effect \citep{OV,Sunyaev:1980nv}, owing to homogeneously ionized gas, is suppressed at these high redshifts, due to relatively low gas temperatures and small density contrasts, the inhomogeneity of ionized hydrogen adds a large contrast to the signal. The effect has been modeled in various ways \cite{Knox:1998fp,Hu:1999vq,Santos:2003jb, Zahn:2005fn, Mcquinn:2005ce}. In \cite{Zahn:2005fn, Mcquinn:2005ce} the signal was found to peak at $\ell=2000$, where it is still almost two orders of magnitude below the primary CMB signal. It is approximately one order of magnitude below the primary fluctuations at $\ell=3000$, and becomes the dominant contribution over the lensed primary CMB at $\ell=3500-4000$, depending on the details of the reionization model and on how well the thermal Sunyaev-Zel'dovich effect \cite{Zeldovich:1969ff} can be removed from the data (see Figure \ref{fig:cmb_contributions}). The height of the ``patchy reionization bump'' (to be contrasted with the large scale bump in {\em polarization} discussed in the last section) depends on the total extent of the epoch, and its width depends on how much time the ionized fraction spends in various phases. If, for example, reionization stalls in its late stages, there will be relatively more large bubbles. 

\begin{figure*}
\includegraphics[width=12cm,angle=-90]{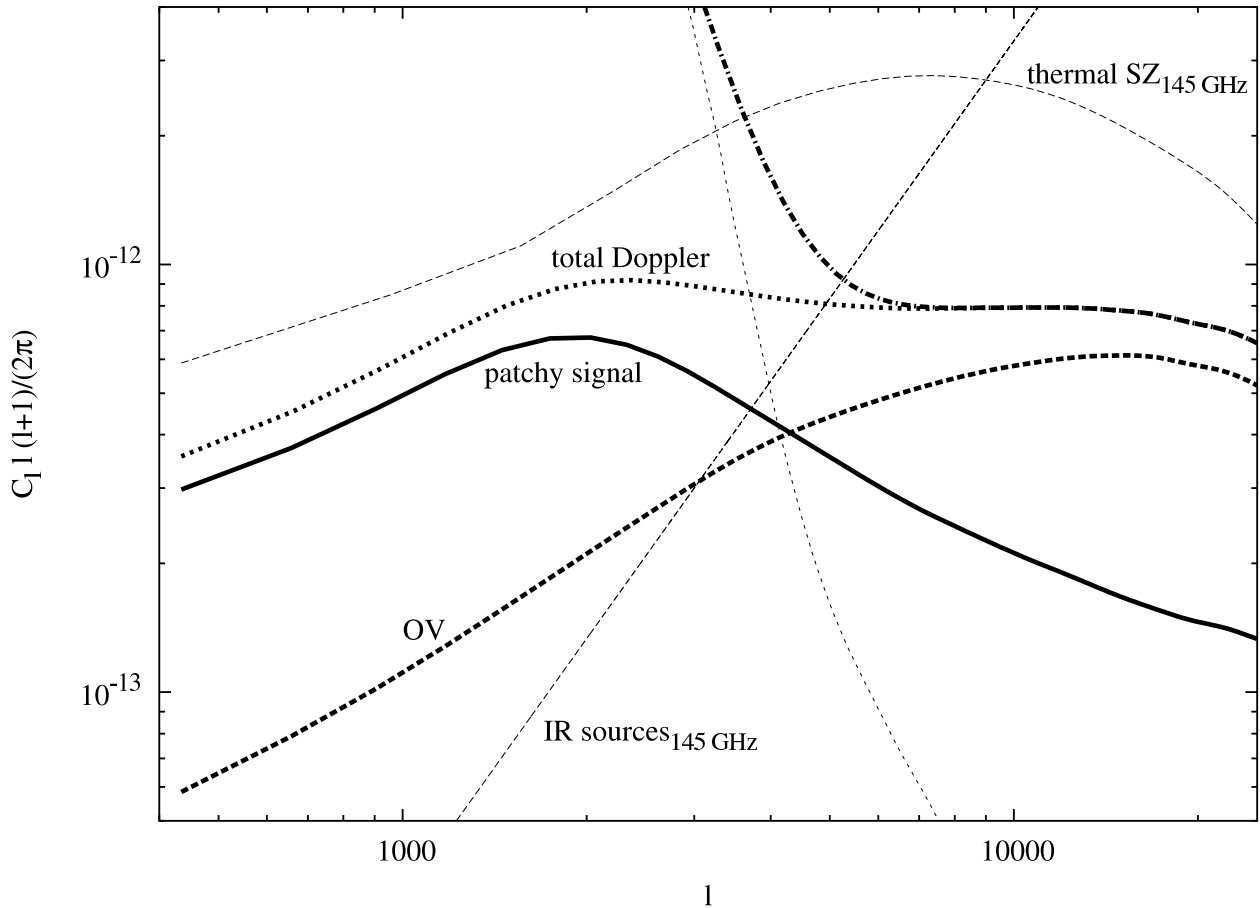}
\caption{Impact of patchy reionization on the small scale CMB anisotropies.
The thick lines show contributions to the small scale CMB power spectrum assuming the spectrally distinguishable signal of thermal SZ and point sources can be ``cleaned''. The patchy reionization excess over regular kSZ peaks at scales $\ell \simeq 2000$. On scales smaller than $\ell \simeq 4000$, the primary CMB vanishes; however, cluster gas physics will have to be reliably modeled in order to discern the patchy reionization contribution from low redshift sources.}
\label{fig:cmb_contributions} 
\end{figure*}

In \cite{Zahn:2005fn} it was shown that already at  Planck sensitivity patchy reionization can bias inferred cosmological parameter values if it is ignored, in particular if only temperature data are used. 
In polarization, secondary anisotropies are expected to be suppressed (see e.g. \cite{Sazonov:1999zp,Hu:1999vq}) by 4 orders of magnitude compared to the primary component. We make the approximation that the patchy reionization signal is unpolarized. The reason temperature data can be biased is that the broad bump seen in figure \ref{fig:cmb_contributions} enhances the power on small angular scales and, if not taken into account in the modeling, leads to (for example) an inferred primordial fluctuation spectrum that is bluer than it really is. The bias can be avoided either by only measuring temperature fluctuations out to $\ell=1500$, or by including an extra parameter for reionization to marginalize over, hoping that this parameter will not degrade the constraints on other parameters much. We go the second route here, and investigate parameter constraints possible with full CMBPol temperature and polarization information. 

In this calculation we assume a raw detector noise level of $w_T^{-1/2}=1 \mu$K-arcmin and $\theta_{\rm FWHM}=3'$. We model reionization following the fast semi-analytic scheme of \cite{Furlanetto:2004nh,Zahn:2005fn}. Here we applied this model in a 1 comoving Gpc/h simulation volume with $1,200^3$ volume elements. We also go beyond previous work in that we vary the threshold for overdensities to be ionized throughout the box while translating between comoving position and redshift, thereby creating a model of ``reionization on the light cone''. The redshift extent of the simulation box covers the entire reionization process. Modeling such a large volume in a full radiative transfer simulation would be prohibitive because of the requirement of a large dynamical range to model both the sources and large scales of reionization. However the size is particularly important for modeling the kinetic SZ effect because the velocity power spectrum peaks on scales of ~250 Mpc/h. Our fiducial model, which will be used as template in the Fisher matrix analysis below, has a value for the ionization efficiency of 15, which leads in our concordance cosmology to an ionization fraction of 0.5 at z=11. To obtain a template for the patchy kSZ signal, we integrate the kinetic SZ signal 
\beq
\frac{\Delta T_{\rm kSZ}}{T_{\rm CMB}} (\n) = - \frac{\sigma_T}{c} \overline{n}_p (\eta\
_0) \int d
\eta [a^{-2} e^{-\tau_{ri}(\eta)} \bar x_e(\eta)] \n \cdot q
\eeq
with
\beq
q=(1+\frac{\delta x_e}{\bar x_e(\eta)}) (1+\delta \omega_{\rm b}) \mathbf{v}
\, .
\eeq
where the Thomson scattering optical depth is $\sigma_T$, $\tau_{ri}$ is the optical 
depth from the observer to conformal time $\eta$, and $\n$ is the line of sight unit vector.

Assuming that the total duration of the EoR is long compared to the scale corresponding to the peak of the velocity power spectrum, we can make the approximation that if reionization takes e.g. twice as long to proceed, the r.m.s. of the induced anisotropy will be $\sqrt{2}$ larger and the power spectrum amplitude doubled. Therefore to first order a 10\% constraint on the amplitude of the patchy reionization admixture to the total CMB power spectrum can be translated into a 10\% constraint on the duration of the process. 

We assume here that the thermal SZ component, which is roughly comparable to the kinetic component on large scales (see Fig. \ref{fig:cmb_contributions}), can be cleaned sufficiently. This can be justified by the fact that at $\ell \simeq 2000$, which is where the patchy reionization contribution peaks, most of the thermal SZ signal stems from massive and hot clusters of mass $M \sim 5 \times 10^{14} \mpch$ and larger at low redshift (see e.g. Figures 5 and 6 of \cite{Komatsu:2002wc}). These large bright objects can hopefully be excised using the multi-frequency information of the experiment \footnote{The brightness of polarized foregrounds will make observations in multiple frequency bands, roughly covering the range most interesting for the SZ, a necessity for a future CMB mission aiming to study the primordial E and B mode signals.}, as well as X-ray data from ROSAT or planned missions like the Constellation-X or X-ray Evolving Universe Spectroscopy (XEUS) missions \footnote{For details on planned X-ray observato- 
ries see http://constellation.gsfc.nasa.gov/ and 
http://www.rssd.esa.int/index.php?project=XEUS.}. Furthermore, the thermal SZ vanishes at 220 GHz, leaving an unobstructed view on the patchy reionization contribution to an experiment observing in this frequency band.

Table \ref{tab:fisher_constraints_patchy} shows  constraints that can be obtained on 8 cosmological parameters including a running of the spectral index $\alpha_s$ and the patchy reionization amplitude. One can see that Planck will not be able to place interesting constraints on the reionization model. In \cite{Zahn:2005fn} an extended reionization model (their model C) was investigated which was able to bias Planck constraints more strongly and might be borderline detectable. 

In the case of CMBPol we find most parameter biases to be large (first column), since they are all  much more tightly constrained. When the patchy reionization ``nuisance parameter'' is included in the analysis, we see that CMBPol should be able to place sub-percent constraints on this amplitude parameter, see table \ref{tab:fisher_constraints_patchy}.

Here we have not discussed how well CMBPol might be able to discern reionization models with different source properties. For example feedback processes \cite{Furlanetto:2005ax} or recombinations \cite{Furlanetto:2005xx} could lead to relatively larger/smaller typical bubble sizes, imprinting the characteristic bump in Figure \ref{fig:cmb_contributions} on larger/smaller scales (and changing the slope at high $\ell$, where the reionization signal comes to dominate over the primordial CMB). Furthermore, the width of the patchy reionization ``bump'' will depend on how much time is spent in each partial ionization stage. 

A number of ground based observatories with high angular resolution, in particular the South Pole Telescope (SPT)\footnote{The website for SPT is http://spt.uchicago.edu/}
and the Atacama Cosmology Telescope (ACT)\footnote{http://physics.princeton.edu/act} are presently beginning observations. These experiments will primarily target the CMB anisotropies on the scale of a few arcminutes, to constrain the thermal SZ and the kinetic SZ  effects from clusters and large scale structure at lower redshift as well as gravitational lensing. They also may be able to detect the patchy reionization admixture on these scales. 

\begin{table}
\bc
\begin{tabular}{|c|c|c|c|}
\hline
$P$ [fid. value] &  $\frac{\Delta P}{P} (\%) $, RefExp (Planck) &  RefExp, $l_{\rm max}=3000$ & RefExp, $l_{\rm max}=2000$ \\
\hline
$\tau_{ri} [0.084]$ &  2.70 (6.97)               &   2.73   & 2.78 \\
$\Omega_\Lambda [0.742] $ & 0.30 (0.89)  & 0.31   & 0.41 \\
$\omega_{dm} [0.1093] $ &  0.41 (1.15)     &  0.44 & 0. 56 \\ 
$\omega_{b} [0.02255] $ & 0.11 (0.68)       & 0.12   & 0.23\\
$n_s$ [0.963] & 0.17 (0.47)                         & 0.18  & 0.21 \\
$A_s$ [$2.4 \, 10^{-9}$] & 0.38 (1.22)            &  0.48  & 0.52 \\
$\alpha_s [0] $ & 0.20 (0.60)                          & 0.23 & 0.33 \\
$A_{PR} [1]$ & 0.42 (119.0)                             & 3.22 & 33.5 \\
\hline
\end{tabular}
\ec
\caption{Forecasted constraints, for a full sky Reference Experiment with noise level $1\mu {\rm K-arcmin}$ ($\sqrt{2} \mu$K-arcmin in polarization) and a $1'$ FWHM beam and Planck from combined temperature and polarization information on 8 cosmological parameters including the patchy reionization amplitude parameter $A_{\rm PR}$, which crudely parametrizes the duration of the EoR. Also included are the results for the reference experiment for small scale cutoffs, which could be due to limited telescope/beam size or point source confusion. An $l_{\rm max}$ of 3000(2000) would correspond to a 1.5(1)m telescope primary at 220 GHz.}           
\label{tab:fisher_constraints_patchy}
\end{table}

\subsection{The effect of metals}

As the first sources turn on an ionize the universe metals are also produced for the first time. These metals  eventually enrich the inter-galactic medium (e.g. \cite{BarkanaLoeb2001}).
Here we concentrate on the resonant scattering signature of metal enrichment of the inter-galactic medium by the  first stars  on the  CMB.
CMB photons interacting with a species $X$ at redshift $z_{rs}$ via a
resonant transition of resonant frequency $\nu_{rs}$, show the effect of the interaction if
$z_{rs} = \nu_{rs}/\nu_{obs} -1$ \citep{basu}.
The resonant
scattering of this species generates an optical depth to CMB
photons  \citep{sobolev}: $\tau_{X} (z) \propto  f_{rs}  n_X(z),$ where  $n_{X}$ is the number density of the $X$ species and $f_{rs}$ is the
absorption oscillator strength. $\tau_X$ modifies temperature anisotropies:
$\Delta_T  = e^{-\tau_{X}} \Delta_{T_{orig}}
        + \Delta_{T_{new}}\,. $
In the limit of  low metal abundance $\tau_X \ll 1$, we can retain
only linear terms in $\tau_X$.
Thus as described in \citep{basu}, the change in the anisotropies
($\Delta_T - \Delta_{T_{orig}}$) reduces to a {\em blurring} factor
($-\tau_X \Delta_{T_{orig}}$) plus a new term
which is important only in the large scales. The large scale signal  
is less clean to interpret than the high $\ell$ signal because at  low multipoles generation and suppression terms tend to cancel out.
The high $\ell$ signal in multipole space is \citep{basu}:
$\Delta a_{\ell m} = - \tau_X a_{\ell m}^{CMB}$, i.e. it is proportional to the primary anisotropies, with  a well defined frequency dependence:
in the presence of only one resonant species at  $z_{rs}$ the effect will be non-zero only at frequency  $\nu_o=\nu_{rs}/(1+z_{rs})$.
Thus  the metals  contribution
to the power spectrum  is proportional to the primordial CMB power
spectrum with constant of proportionality given by $\tau_{\nu_{rs}}$.

In a given experiment, the lowest frequency channel corresponds to the highest redshift where the metal abundance is the lowest.  Increments of metal abundances  can be measured by studying
the difference map and the difference power spectra between channels.
Signal-to-noise considerations  \citep{HMVJ} show that the main limiting factor  is the accuracy of the cross-channel calibration.

A purpose-built experiment with the properties of the EPIC 2m set up, with accurate control of systematics and cross-channel calibration might be sensitive to changes in 3-12\% (2-$\sigma$) of the Solar fraction of Oxygen abundance at $12<z<22$  for reionizaton redshift $z_{re}>10$ (OIII $84$ $\mu$ m transition) and to  changes in N abundance of 60\% its solar value for $5.5<z<9$ (2-$\sigma$)(NII 205 $\mu$m  transition).



\section*{Appendix: Likelihood Function}
For computing the constraints from the WMAP 5-year data, we shall evaluate the
likelihood function directly in pixel space, following Appendix D of
Page et al. (2007). 

For forecasting parameter constraints from Planck and CMBPol, we
shall evaluate the likelihood functions in harmonic space. 
The form of the likelihood function for a theoretical model, $C_l$, given the
data, $C_l^{data}$, is 
\begin{eqnarray}
\nonumber
 -2\ln L(C_l|C_l^{data})
&=& f_{sky}\sum_{l=2}^{l_{max}} 
(2l+1)
\left\{
\ln\left[C_l^{EE}C_l^{TT}-(C_l^{TE})^2\right]\right.\\
& &\left.+\frac{C_l^{EE}(C_l^{TT,data}+N_l^{TT})+C_l^{TT}(C_l^{EE,data}+N_l^{EE})-2C_l^{TE}C_l^{TE,data}}{C_l^{EE}C_l^{TT}-(C_l^{TE})^2}
\right\}, 
\end{eqnarray}
where $N_l$ are the noisebias (which is assumed to be constant over
multipoles) given by
\begin{eqnarray}
 N_l^{TT} &=& \left(\frac{\pi}{10800}\frac{w_t^{-1/2}}{\mu
	       K~{\rm arcmin}}\right)^2~\mu K^2~{\rm str},\\
 N_l^{EE} &=&  \left(\frac{\pi}{10800}\frac{w_p^{-1/2}}{\mu
		K~{\rm arcmin}}\right)^2~\mu K^2~{\rm str}.
\end{eqnarray}
Here, $w_t^{-1/2}$ and $w_p^{-1/2}$ are the temperature and polarization
sensitivities, respectively, which are related to ``$\Delta T/T$ per
pixel'' in Table 1.1 on pp.4 of the Planck Blue Book via
\begin{eqnarray}
\nonumber
 w_t^{-1/2} &=& 2.725~K\times (\mbox{$\Delta T/T$ per pixel for Stokes $I$
  in units of
  $\mu K/K$})\\ & &\times (\mbox{Angular Resolution in {\rm arcmin}}),\\
\nonumber
 w_p^{-1/2} &=& 2.725~K\times (\mbox{$\Delta T/T$ per pixel for Stokes
  $Q$ \& $U$
  in units of
  $\mu K/K$})\\ & &\times (\mbox{Angular Resolution in {\rm arcmin}}).
\end{eqnarray}
Note that we have ignored the effect of beam smearing, as we focus only
on low multipoles where the beam effect may be ignored safely.

In our calculation we integrate the likelihood function up to
$l_{max}=30$, beyond which little information is available for
reionization. When we compute the likelihood function for CMBPol, we
shall assume that 
CMBPol is limited only by the cosmic variance up to $l_{max}$, i.e.,
$N_l^{TT}=0=N_l^{EE}$ up to $l_{max}=30$. On the other hand when we
compute the likelihood function for Planck, we shall combine the
sensitivities in 70~GHz (LFI), 100~GHz (HFI), and 143~GHz (HFI), such
that 
\begin{eqnarray}
 w_t^{-1/2} &=& \frac1{\sqrt{(w^{-1/2}_{t,70~GHz})^{-2} 
+ (w^{-1/2}_{t,100~GHz})^{-2} + (w^{-1/2}_{t,143~GHz})^{-2}}},\\
w_p^{-1/2} &=& \frac1{\sqrt{(w^{-1/2}_{p,70~GHz})^{-2}
+ (w^{-1/2}_{p,100~GHz})^{-2} + (w^{-1/2}_{p,143~GHz})^{-2}}}.
\end{eqnarray}
From the Planck Blue Book we find
$(w^{-1/2}_{t,70~GHz},w^{-1/2}_{t,100~GHz},w^{-1/2}_{t,143~GHz})=(179.3,68.1,42.6)~\mu
K~{\rm arcmin}$, and $(w^{-1/2}_{p,70~GHz},w^{-1/2}_{p,100~GHz},w^{-1/2}_{p,143~GHz})=(255.6,109.0,81.3)~\mu
K~{\rm arcmin}$. Therefore, the combined sensitivities for Planck are given by
\begin{eqnarray}
 w_t^{-1/2} &=& 35.4~\mu K~{\rm arcmin},\\
 w_p^{-1/2} &=& 63.1~\mu K~{\rm arcmin},
\end{eqnarray}
and thus we find
\begin{eqnarray}
 N_l^{TT} &=& 1.06\times 10^{-4}~\mu K^2~{\rm str},\\
 N_l^{EE} &=& 3.37\times 10^{-4}~\mu K^2~{\rm str}.
\end{eqnarray}

\newpage

\bibliographystyle{h-physrev3.bst}

\begin{thebibliography}{99}

\bibitem{Komatsu08}
  E.~Komatsu {\it et al.}  [WMAP Collaboration],
  arXiv:0803.0547 [astro-ph].


\bibitem{wyithe03}
{Wyithe}, J.~S.~B., \& {Loeb}, A. 2003, \apjl, 588, L69

\bibitem{cen03}
{Cen}, R. 2003, \apjl, 591, L5

\bibitem{haiman03}
{Haiman}, Z., \& {Holder}, G.~P. 2003, \apj, 595, 1

\bibitem{mackey03}
{Mackey}, J., {Bromm}, V., \& {Hernquist}, L. 2003, \apj, 586, 1

\bibitem{yoshida03-semian}
{Yoshida}, N., {Abel}, T., {Hernquist}, L., \& {Sugiyama}, N. 2003, \apj, 592,
  645

\bibitem{ybh}
{Yoshida}, N., {Bromm}, V., \& {Hernquist}, L. 2004, \apj, 605, 579

\bibitem{sok03}
Sokasian, A., Abel, T., Hernquist, L. \&
Springel, V. 2003, MNRAS, 344, 607

\bibitem{sok04}
Sokasian, A., Yoshida, N., Abel, T., Hernquist, L. \&
Springel, V. 2004, MNRAS, 350, 47

\bibitem{yoshida06}
Yoshida, N., Omukai, K., Hernquist, L. \& Abel, T. 2006,
\apj, 652, 6

\bibitem{zal97}
Zaldarriaga, M.\ 1997,  \prd, 55, 1822 

\bibitem{kaplinghat03}
{Kaplinghat}, M., {Chu}, M., {Haiman}, Z., {Holder}, G.~P., {Knox}, L., \&
  {Skordis}, C. 2003, \apj, 583, 24


\bibitem{hu03}
{Hu}, W., \& {Holder}, G.~P. 2003, \prd, 68, 23001

\bibitem{McQuinn:2006et} 
M.~McQuinn, A.~Lidz, O.~Zahn, S.~Dutta, L.~Hernquist \& M.~Zaldarriaga
%
Mon.\ Not.\ Roy.\ Astron.\ Soc.\  {\bf 377}, 1043 (2007) [arXiv:astro-ph/0610094].

\bibitem{Zahn:2006sg}
  O.~Zahn, A.~Lidz, M.~McQuinn, S.~Dutta, L.~Hernquist, M.~Zaldarriaga \& S.~R.~Furlanetto,
  Astrophys.\ J.\  {\bf 654}, 12 (2006)
  [arXiv:astro-ph/0604177].


%

%


%



\bibitem{becker}
{Becker}, R.~H., {et~al.} 2001, \aj, 122, 2850

\bibitem{Fan:2005es}
  X.~H.~Fan {\it et al.},
  Astron.\ J.\  {\bf 132}, 117 (2006)
  [arXiv:astro-ph/0512082].

\bibitem{Songaila04} Songaila, A. 2004, AJ, 127, 2598

\bibitem{Becker:2006qj}
Becker, G.~D., Rauch, M., \& Sargent, W.~L.~W. 2006, astro-ph/06070633

\bibitem{Miralda00} Miralda-Escud\'e, J., Haehnelt, M., \& Rees, M.~J. 2000, ApJ, 530, 1

\bibitem{Oh05} Oh, S.~P., \& Furlanetto, S.~R., 2005, ApJL, 620, 9

\bibitem{Lidz06a} Lidz, A., Oh, S.~P., \& Furlanetto, S.~R. 2006, ApJL, 639, 47

\bibitem{Wyithe:2005sy} Wyithe, J.~S.~B., \& Loeb, A. 2006, ApJ, 646, 696

\bibitem{Liu06} Liu, J., Bi, H., Feng, L.~L., \& Fang, L.~Z., 2006, ApJL, 645, 1


\bibitem{Wyithe:2004mx}
J.~S.~B.~Wyithe \& A.~Loeb 2004,
arXiv:astro-ph/0401188.

\bibitem{Wyithe:2004jw}
  J.~S.~B.~Wyithe, A.~Loeb \& C.~Carilli,
  Astrophys.\ J.\  {\bf 628}, 575 (2005)
  [arXiv:astro-ph/0411625].

\bibitem{Mesinger:2006kn}
  A.~Mesinger \& Z.~Haiman,
  arXiv:astro-ph/0610258.

\bibitem{Bolton:2006pc}
  J.~S.~Bolton \& M.~G.~Haehnelt,
  Mon.\ Not.\ Roy.\ Astron.\ Soc.\  {\bf 374}, 493 (2007)
  [arXiv:astro-ph/0607331].


\bibitem{Lidz:2007mz}
  A.~Lidz, M.~McQuinn \& M.~Zaldarriaga
  
Astrophys.\ J.\  {\bf 670}, 39 (2007) [arXiv:astro-ph/0703667].
%


\bibitem{Ryan-Weber:2006}
Ryan-Webber, E.~V., Pettini, M., \& Madau, P., 2006, MNRAS, 371L, 78

\bibitem{Simcoe:2006}
Simcoe, R.~A., 2006, ApJ, 653, 977

\bibitem{Oh:2002} 
Oh, S.~P. 2002, MNRAS, 336, 1021

\bibitem{Becker:2006b} 
Becker, G.~D., Sargent, W.~L.~W., Rauch, M., 
\& Simcoe, R.~A. 2006, ApJ, 640, 69

\bibitem{theuns02-reion}
{Theuns}, T., {et~al.} 2002, \apjl, 567, L103

\bibitem{hui03}
{Hui}, L., \& {Haiman}, Z. 2003, \apj, 596, 9

\bibitem{sokasian02}
{Sokasian}, A., {Abel}, T., \& {Hernquist}, L. 2002, \mnras, 332, 601

\bibitem{McQuinn:2008am}
McQuinn, M., Lidz, A., Zaldarriaga, M., Hernquist, L., Hopkins, P.~F., Dutta, S., 
\& Faucher-Giguere, C.~A., 2008, arXiv:0807.2799

\bibitem{Kohler:2005gg}
Kohler, K., Gnedin, N.~Y., \& Hamilton, A.~J.~S., 2007, ApJ, 657, 15

\bibitem{Trac:2008yz}
Trac, H., Cen, R., \& Loeb, A. 2008, ApJL submitted, arXiv:0807.4530

\bibitem{Partridge67} Partridge, R.~B., \& Peebles, P.~J.~E. 1967, ApJ, 147, 868

\bibitem{hue02}
{Hu}, E.~M., {et~al.} 2002, \apjl, 568, L75

\bibitem{kodaira03}
Kodaira, K., et al.\ 2003, \pasj, 55, 232

\bibitem{rhoads04}
{Rhoads}, J.~E., {et~al.} 2004, \apj, in press (astro-ph/0403161)

\bibitem{stanway04}
{Stanway}, E.~R., {et~al.} 2004, \apjl, 604, L13

\bibitem{santos04-obs}
{Santos}, M.~R., {et~al.} 2004, \apj, 606, 683

\bibitem{kashikawa06}
{Kashikawa} N.,  et~al., 2006, \apj, 648, 7


\bibitem{cuby06}
{Cuby} J.-G.,  {Hibon} P.,  {Lidman} C.,  {Le F{\`e}vre} O.,  {Gilmozzi} R.,
  {Moorwood} A.,    {van der Werf} P.,  2007, \aap, 461, 911

\bibitem{willis05}
{Willis} J.~P.,  {Courbin} F.,  2005, \mnras, 357, 1348


\bibitem{stark07a}
  D.~P.~Stark, R.~S.~Ellis, J.~Richard, J.~P.~Kneib, G.~P.~Smith \& M.~R.~Santos,
  arXiv:astro-ph/0701279.


\bibitem{casali06}
{Casali} M.,  et~al., 2006, in Ground-based and Airborne Instrumentation for
  Astronomy. Edited by McLean, Ian S.; Iye, Masanori. Proceedings of the SPIE,
  Volume 6269, pp. (2006). {HAWK-I: the new wide-field IR imager for the VLT}

\bibitem{mcpherson06}
{McPherson} A.~M.,  et~al., 2006, in Ground-based and Airborne Telescopes.
  Edited by Stepp, Larry M.. Proceedings of the SPIE, Volume 6267, pp. (2006).
  {VISTA: project status}

\bibitem{horton04}
Horton A.,  Parry I.,  Bland-Hawthorn J.,  Cianci S.,  King D.,  McMahon R.,
  Medlen S.,  2004, PROC.SPIE INT.SOC.OPT.ENG., 5492, 1022


\bibitem{McQuinn07b} McQuinn, M., Hernquist, L., Zaldarriaga, M., \& Dutta, S. 2007, MNRAS, 381, 75

\bibitem{Miralda98} Miralda-Escud\'e, J. 1998, ApJ, 501, 15

\bibitem{Furlanetto06c} Furlanetto, S.~R., Zaldarriaga, M., \& Hernquist, L. 2006, MNRAS, 365, 1012

\bibitem{Mesinger07b} Mesinger, A., \& Furlanetto, S.~R., 2007, MNRAS in press, arXiv:0708.0006

\bibitem{Malhotra06} Malhotra, S. \& Rhoads, J.~E., 2006, ApJL, 647, 95

\bibitem{Dijkstra07} Dijkstra, M., Wyithe, J.~S.~B., \& Haiman, Z., 2007,
MNRAS, 379, 253

\bibitem{Fernandez08} Fernandez, E.~R., \& Komatsu, E., 2008, 
MNRAS, 384, 1363

\bibitem{Barkana04} Barkana, R., \& Loeb, A., 2004, ApJ, 601, 64

\bibitem{Totani:2006} Totani, T., Kawai, N., Kosugi, G., Aoki, K., Yamada, T., Iye, M.,
Ohta, K., \& Hattori, T., 2006, PASJ, 58, 485


\bibitem{McQuinn:2007gm}
  M.~McQuinn, A.~Lidz, M.~Zaldarriaga, L.~Hernquist \& S.~Dutta,
  arXiv:0710.1018 [astro-ph].

\bibitem{Chen:2007} Chen, H.~-W., Prochaska, J.~X., \& Gnedin, N.~Y. 2007, ApJL, 667, 125

\bibitem{chen03}
X.~L.~Chen \& J.~Miralda-Escud\'e 2004,
Astrophys.\ J.\  602, 1


\bibitem{ciardi03}
Ciardi, B., \& Madau, P., 2003, ApJ, 596, 1

\bibitem{Furlanetto:2004nh}
  S.~Furlanetto, M.~Zaldarriaga \& L.~Hernquist,
  Astrophys.\ J.\  {\bf 613}, 1 (2004)
  [arXiv:astro-ph/0403697].

\bibitem{Iliev08} Iliev, I.~T., Mellema, G., Pen, U.~L., Bond, J.~R.,
\& Shapiro, P.~R. 2008, MNRAS, 384, 863

\bibitem{Furlanetto06a} Furlanetto, S.~R., Oh, S.~P., \& Briggs, F. 2006, Phys. Reports, 433, 181


\bibitem{McQuinn:2005hk}
  M.~McQuinn, O.~Zahn, M.~Zaldarriaga, L.~Hernquist \& S.~R.~Furlanetto,
  Astrophys.\ J.\  {\bf 653}, 815 (2006)
  [arXiv:astro-ph/0512263].

\bibitem{Lidz:2007az}
  A.~Lidz, O.~Zahn, M.~McQuinn, M.~Zaldarriaga \& L.~Hernquist,
  arXiv:0711.4373 [astro-ph].
  
  \bibitem{Pritchard07} Pritchard, J.~R., \& Furlanetto, S.~R. 2007, MNRAS, 376, 1680

\bibitem{Zaldarriaga:2003du}
M.~Zaldarriaga, S.~R.~Furlanetto \& L.~Hernquist 2004,
\apj, 608, 622

\bibitem{Dunetal08}
  J.~Dunkley {\it et al.}  [WMAP Collaboration],
  arXiv:0803.0586 [astro-ph].

\bibitem{ColPierPri08}
 L.~P.~L.~Colombo {\it e; al.} In preparation.  

\bibitem{MorHu08c}
  M.~J.~Mortonson \& W.~Hu,
 ApJ Lett., 686, L53 (2008) [arXiv:0804.2631].


\bibitem{ColPie08}
  L.~P.~L.~Colombo \& E.~Pierpaoli,
  arXiv:0804.0278 [astro-ph].

\bibitem{Crill08}
  B.~P.~Crill {\it et al.},
  arXiv:0807.1548 [astro-ph].


\bibitem{EPICpap}
  J.~Bock {\it et al.},
  arXiv:0805.4207 [astro-ph].

\bibitem{MorHu08a}
  M.~J.~Mortonson \& W.~Hu,
  Astrophys.\ J.\  {\bf 672}, 737 (2008)
  [arXiv:0705.1132 [astro-ph]].

\bibitem{Alvarez06} 
Alvarez, M.~A., Komatsu, E., Dor\'e, O., \& Shapiro, P.~R. 2006, ApJ, 647, 840

\bibitem{FurLoe05}
  S.~Furlanetto \& A.~Loeb,
  Astrophys.\ J.\  {\bf 634}, 1 (2005)
  [arXiv:astro-ph/0409656].

\bibitem{LewWelBat06}
  A.~Lewis, J.~Weller \& R.~Battye,
  Mon.\ Not.\ Roy.\ Astron.\ Soc.\  {\bf 373}, 561 (2006)
  [arXiv:astro-ph/0606552].



\bibitem{HuHol03}
{Hu}, W., \& {Holder}, G.~P. 2003, \prd, 68, 23001


\bibitem{MLidd08}
  P.~Mukherjee \& A.~R.~Liddle,
  arXiv:0803.1738 [astro-ph].

\bibitem{MorHu08b}
  M.~J.~Mortonson \& W.~Hu,
  Phys.\ Rev.\  D {\bf 77}, 043506 (2008)
  [arXiv:0710.4162 [astro-ph]].


\bibitem[{Ostriker \& Vishniac(1986)}]{OV}
Ostriker, J. \& Vishniac, E. 1986, \apj, 306, 51


\bibitem[{Sunyaev \& Zel'dovich(1980)}]{Sunyaev:1980nv}
Sunyaev, R.~A. \& Zel'dovich, Y.~B. 1980, MNRAS, 190, 413


\bibitem[{Knox {et~al.}(1998)Knox, Scoccimarro, \& Dodelson}]{Knox:1998fp}
Knox, L., Scoccimarro, R., \& Dodelson, S. 1998, Phys. Rev. Lett., 81, 2004

\bibitem{Hu:1999vq}
  W.~Hu,
    Astrophys.\ J.\  {\bf 529}, 12 (2000)
  [arXiv:astro-ph/9907103].
  
\bibitem[{Santos {et~al.}(2003)Santos, Cooray, Haiman, Knox, \&
  Ma}]{Santos:2003jb}
Santos, M.~G., Cooray, A., Haiman, Z., Knox, L., \& Ma, C.-P. 2003,
\apj, 598, 756

\bibitem[{Zahn {et~al.}(2005)Zahn, Zaldarriaga, Hernquist, \&
  McQuinn}]{Zahn:2005fn}
Zahn, O., Zaldarriaga, M., Hernquist, L., \& McQuinn, M. 2005, ApJ, 630, 657,
  \eprint{astro-ph/0503166}
  
  \bibitem[{McQuinn {et~al.}(2005)McQuinn, Furlanetto, Hernquist, Zahn, \&
  Zaldarriaga}]{Mcquinn:2005ce}
McQuinn, M., Furlanetto, S.~R., Hernquist, L., Zahn, O., \& Zaldarriaga, M.
  2005, ApJ, 630, 643, \eprint{astro-ph/0504189}
  
\bibitem[{Zel'dovich \& Sunyaev(1969)}]{Zeldovich:1969ff}
Zel'dovich, Y.~B. \& Sunyaev, R.~A. 1969, Astrophys. Space Sci., 4, 301


\bibitem{Sazonov:1999zp}
  S.~Y.~Sazonov \& R.~A.~Sunyaev,
  Mon.\ Not.\ Roy.\ Astron.\ Soc.\  {\bf 310}, 765 (1999)
  [arXiv:astro-ph/9903287].

 
\bibitem{Komatsu:2002wc}
  E.~Komatsu \& U.~Seljak,
  Mon.\ Not.\ Roy.\ Astron.\ Soc.\  {\bf 336}, 1256 (2002)
  [arXiv:astro-ph/0205468].
  

  \bibitem[{Furlanetto {et~al.}(2006{\natexlab{a}})Furlanetto, McQuinn, \&
  Hernquist}]{Furlanetto:2005ax}
Furlanetto, S.~R., McQuinn, M., \& Hernquist, L. 2006{\natexlab{a}}, MNRAS,
  365, 115, \eprint{astro-ph/0507524}
  
  \bibitem[{Furlanetto \& Oh(2005)}]{Furlanetto:2005xx}
Furlanetto, S.~R., \& Oh, S.~P. 2005, MNRAS, 363, 103, \eprint{astro-ph/0505065\
}




\bibitem{BarkanaLoeb2001} Barkana R., Loeb A., 2001, PhR, 349, 125
\bibitem{basu} Basu, K.,
Hern{\'a}ndez-Monteagudo, C., \& Sunyaev, R.~A.\ 2004, A\& A, 416, 447
\bibitem{sobolev} Sobolev, V.V. 1946, Moving Atmospheres of
Stars (Lenningrad:
Leningrad State Univ.; English transl.1960, Cambridge:Harvard
Univ. Press)
\bibitem{HMVJ} Hern\'andez-Monteagudo, C., Verde, L., Jimenez, R.  \ 2006,  ApJ, 653, 10.

\end{thebibliography}

\end{document}